\documentclass[review]{elsarticle}

\usepackage{graphicx}	% Include figure files
\usepackage{dcolumn}	% Align table columns on decimal point
\usepackage{bm}		% Bold math

\usepackage{amsthm}
\usepackage{amsmath}
\usepackage{units}

\usepackage{times}
\usepackage{epstopdf}

\usepackage{caption}
\DeclareGraphicsExtensions{.pdf}

\begin{document}

\title{The Large Underground Xenon (LUX) Experiment}

\author[cwru]{D.\,S.\,Akerib}

\author[sdsmt]{X.\,Bai}

\author[yale]{S.\,Bedikian}

\author[yale]{E.\,Bernard}

\author[llnl]{A.\,Bernstein}

\author[moscow]{A.\,Bolozdynya}

\author[cwru]{A.\,Bradley}

\author[usd]{D.\,Byram}

\author[yale]{S.\,B.\,Cahn}

\author[tamu]{C.\,Camp}

\author[cwru]{M.\,C.\,Carmona-Benitez}

\author[llnl]{D.\,Carr}

\author[brown]{J.\,J.\,Chapman}

\author[usd]{A.\,Chiller}

\author[usd]{C.\,Chiller}

\author[cwru]{K.\,Clark}

\author[davis]{T.\,Classen}

\author[cwru]{T.\,Coffey}

\author[yale]{A.\,Curioni}

\author[cwru]{E.\,Dahl}

\author[llnl]{S.\,Dazeley}

\author[lip]{L.\,de\,Viveiros}

\author[umd]{A.\,Dobi}

\author[cwru]{E.\,Dragowsky}

\author[uor]{E.\,Druszkiewicz}

\author[yale]{B.\,Edwards}

\author[brown]{C.\,H.\,Faham}

\author[brown]{S.\,Fiorucci}

\author[brown]{R.\,J.\,Gaitskell}

\author[cwru]{K.\,R.\,Gibson}

\author[lbl]{M.\,Gilchriese}

\author[umd]{C.\,Hall}

\author[sdsmt]{M.\,Hanhardt}

\author[davis]{B.\,Holbrook}

\author[ucb]{M.\,Ihm}

\author[ucb]{R.\,G.\,Jacobsen}

\author[yale]{L.\,Kastens}

\author[llnl]{K.\,Kazkaz}

\author[umd]{R.\,Knoche}

\author[ucsb]{S.\,Kyre}

\author[case]{J.\,Kwong}

\author[davis]{R.\,Lander}

\author[yale]{N.\,A.\,Larsen}

\author[cwru]{C.\,Lee}

\author[umd]{D.\,S.\,Leonard}

\author[lbl]{K.\,T.\,Lesko}

\author[lip]{A.\,Lindote}

\author[lip]{M.\,I.\,Lopes}

\author[yale]{A.\,Lyashenko}

\author[brown]{D.\,C.\,Malling}

\author[tamu]{R.\,Mannino}

\author[tamu]{Z.\,Marquez}

\author[yale]{D.\ N.\,McKinsey}

\author[usd]{D.\,-M.\,Mei}

\author[davis]{J.\,Mock}

\author[uor]{M.\,Moongweluwan}

\author[harvard]{M.\,Morii}

\author[ucsb]{H.\,Nelson}

\author[lip]{F.\,Neves}

\author[yale]{J.\,A.\,Nikkel}

\author[brown]{M.\,Pangilinan}

\author[yale]{P.\,D.\,Parker}

\author[yale]{E.\,K.\,Pease}

\author[cwru]{K.\,Pech}

\author[cwru]{P.\,Phelps}

\author[tamu]{A.\,Rodionov}

\author[tamu]{P.\,Roberts}

\author[davis]{A.\,Shei}

\author[cwru]{T.\,Shutt}

\author[lip]{C.\,Silva}

\author[uor]{W.\,Skulski}

\author[lip]{V.\,N.\,Solovov}

\author[tamu]{C.\ J.\, Sofka}

\author[llnl]{P.\,Sorensen}

\author[usd]{J.\,Spaans}

\author[tamu]{T.\,Stiegler}

\author[davis]{D.\,Stolp}

\author[davis]{R.\,Svoboda}

\author[davis]{M.\,Sweany}

\author[davis]{M.\,Szydagis}

\author[sdsta]{D.\,Taylor}

\author[davis]{J.\,Thomson}

\author[davis]{M.\,Tripathi}

\author[davis]{S.\,Uvarov}

\author[brown]{J.\,R.\,Verbus}

\author[davis]{N.\,Walsh}

\author[tamu]{R.\,Webb}

\author[ucsb]{D.\,White}

\author[tamu]{J.\,T.\,White}

\author[cwru]{T.\,J.\,Whitis}

\author[harvard]{M.\,Wlasenko}

\author[uor]{F.\,L.\,H.\,Wolfs \corref{cor1}}
\cortext[cor1]{Corresponding Author: wolfs@pas.rochester.edu}

\author[davis]{M.\,Woods}

\author[usd]{C.\,Zhang}

\address[brown]{Brown University, Dept. of Physics, 182 Hope St., Providence, RI
02912, United States}

\address[cwru]{Case Western Reserve University, Dept. of Physics, 10900 Euclid
Ave, Cleveland, OH 44106, United States}

\address[harvard]{Harvard University, Dept. of Physics, 17 Oxford St., Cambridge,
MA 02138, United States}

\address[lbl]{Lawrence Berkeley National Laboratory, 1 Cyclotron Rd., Berkeley,
CA 94720, United States}

\address[llnl]{Lawrence Livermore National Laboratory, 7000 East Ave., Livermore,
CA 94551, United States}

\address[lip]{LIP-Coimbra, Department of Physics, University of Coimbra, Rua Larga, 3004-516 Coimbra, Portugal}

\address[moscow]{Moscow Engineering Physics Institute, 31 Kashirskoe shosse, Moscow
115409, Russia}

\address[sdsmt]{South Dakota School of Mines and Technology, 501 East St Joseph
St., Rapid City, SD 57701, United States}

\address[sdsta]{South Dakota Science and Technology Authority, Sanford Underground Research Facility, Lead, SD 57754, United States}

\address[tamu]{Texas A \& M University, Dept. of Physics, College Station, TX 77843, United States}

\address[ucb]{University of California Berkeley, Dept. of Physics, Berkeley, CA
94720, United States}

\address[davis]{University of California Davis, Dept. of Physics, One Shields Ave.,
Davis, CA 95616, United States}

\address[ucsb]{University of California Santa Barbara, Dept. of Physics, Santa
Barbara, CA 95616, United States}

\address[umd]{University of Maryland, Dept. of Physics, College Park, MD 20742, United States}

\address[uor]{University of Rochester, Dept. of Physics and Astronomy, Rochester,
NY 14627, United States}

\address[usd]{University of South Dakota, Dept. of Physics, 414E Clark St., Vermillion,
SD 57069, United States}

\address[yale]{Yale University, Dept. of Physics, 217 Prospect St., New Haven,
CT 06511, United States}

\date{\today}  	%  but any date may be explicitly specified

\begin{abstract}
The Large Underground Xenon (LUX) collaboration has designed and constructed a dual-phase xenon detector, in order to conduct a search for Weakly Interacting Massive Particles(WIMPs), a leading dark matter candidate. The goal of the LUX detector is to clearly detect (or exclude) WIMPS with a spin independent cross section per nucleon of $2\times 10^{-46}$~cm$^{2}$, equivalent to $\sim$1~event/100 kg/month in the inner 100-kg fiducial volume (FV) of the 370-kg detector. The overall background goals are set to have $<$1~ background events characterized as possible WIMPs in the FV in 300 days of running. 

This paper describes the design and construction of the LUX detector. 

\end{abstract}

\maketitle

\section{Introduction}
\label{sec:intro}

\begin{figure}[]
\begin{center}
	\includegraphics[width=\columnwidth]{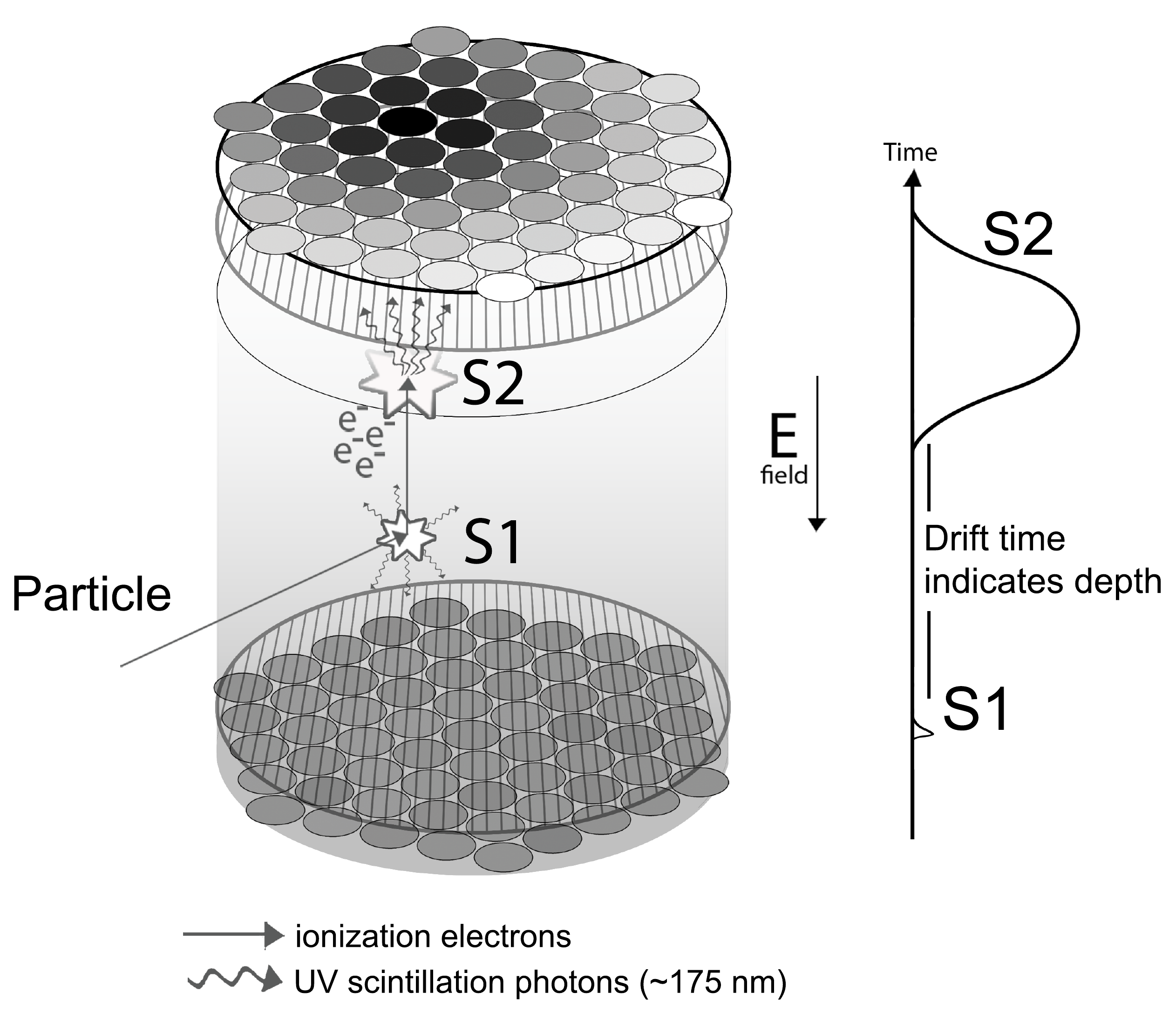} 
	\caption{Schematic of a dark matter interaction in a two-phase xenon detector.  The primary interaction produces scintillation light (S1).  The ionization electrons, created during the primary interaction, create an electro-luminescence pulse (S2) after entering the gas region above the liquid. The S1 and S2 light is captured with two arrays of PMTs, located below and above the active region of the detector.  The intensity of the S2 light captured in each PMT is shown on a gray scale. White and black correspond to the smallest and largest number of photoelectrons detected by the PMTs, respectively. The scattered outgoing dark matter particle is not shown in this Figure.\label{eventFig}}
\end{center}
\end{figure}

Two-phase liquid xenon detectors use a powerful new technology for the detection of dark matter \cite{Akimov2007, Alner2007, Angle2008, Aprile2011, AKimov2012, Aprile2012}.  The operation of a two-phase detector is schematically shown in Fig.~\ref{eventFig}.  Events in the liquid xenon (LXe) target create direct scintillation light (S1), while electrons escaping recombination at the event site are drifted to the liquid surface and extracted into the gas phase by applied electric fields, where they create electro-luminescence light (S2). Both signals are measured by arrays of photomultipliers (PMTs), located above and below the active LXe region. The bottom PMT array largely measures the S1 signal, as photons in the liquid are mostly trapped by total internal reflection.  The top PMT array images the x-y location of the S2 signal, and thus the x-y location of the event site. The drift time, obtained from the time difference between the S1 and S2 signals, gives the depth of the interaction.   This technique thus provides 3D imaging of the event location. 

The noble gas xenon has no naturally occurring radioactive isotopes, and can be readily purified, providing very low internal radioactivity. External backgrounds are minimized by the appropriate choice of detector materials and the design of shielding. Two other methods further reduce the remaining background to levels allowing the detection of WIMPs: (1) discrimination of electron recoils (ER), arising from background $\gamma$ rays and betas, from nuclear recoils (NR) from WIMPs, based on the ratio of charge to scintillation light; and (2) the strong self-shielding capability of the dense LXe, combined with precise 3-D event position determination. 

In this paper we describe the design and construction of the LUX detector which will operate in the Davis Cavern at the Sanford Underground Research Facility (SURF) at Homestake, South Dakota. In Sec.~\ref{sec:design} of this paper, the LUX design is described.  Details of the internals of the LUX detector are discussed in Sec.~\ref{sec:internals}.    The hardware used for detector calibrations is presented in Sec.~\ref{sec:calibrations}.  The purification and gas handling of the xenon is covered in Sec.~\ref{sec:gas}. Light collection and the LUX photomultipliers tubes (PMTs) are the focus of Sec.~\ref{sec:pmts}, and the analog electronics, used to process the PMT signals, and the trigger and data acquisition systems are described in Sec.~\ref{sec:electronics}.  The LUX water shield is discussed in Sec.~\ref{sec:watershield}.  Background studies and measurements are described in Sec.~\ref{sec:bg}.  The integration of the detector on the surface and underground is covered in Sec.~\ref{sec:integration}.

\section{LUX Detector Design and Sensitivity}
\label{sec:design}

The goal of LUX is to detect (or exclude) WIMPS with spin-independent cross sections per nucleon as low as $2\times$10$^{-46}$~cm$^2$, equivalent to $\sim$0.5~events/100 kg/month in the inner 100-kg fiducial volume of the 370-kg LXe detector.  The overall goal is to ensure that during a 10-months running period not a single WIMP candidate in the fiducial volume is due to a background interaction.  The backgrounds consist of electron recoils, primarily from $\gamma$ rays, and nuclear recoils from neutrons.  Based on the calibrations performed in above-ground laboratories \cite{Shutt2007} and data from the XENON10 experiment \cite{Angle2008}, we project that the detector will reject 99.3 - 99.9\% (energy dependent) of electron-recoil events for a 50\% acceptance of nuclear recoil events in the energy range relevant to dark matter detection.  With this rejection power, the goal for the residual low-energy ER background, using appropriate scaling factors for electron and nuclear recoil quenching, can be $>$150 times higher than the rate of background from NR.

\begin{figure}[]
\begin{center}
	\includegraphics[width=\columnwidth]{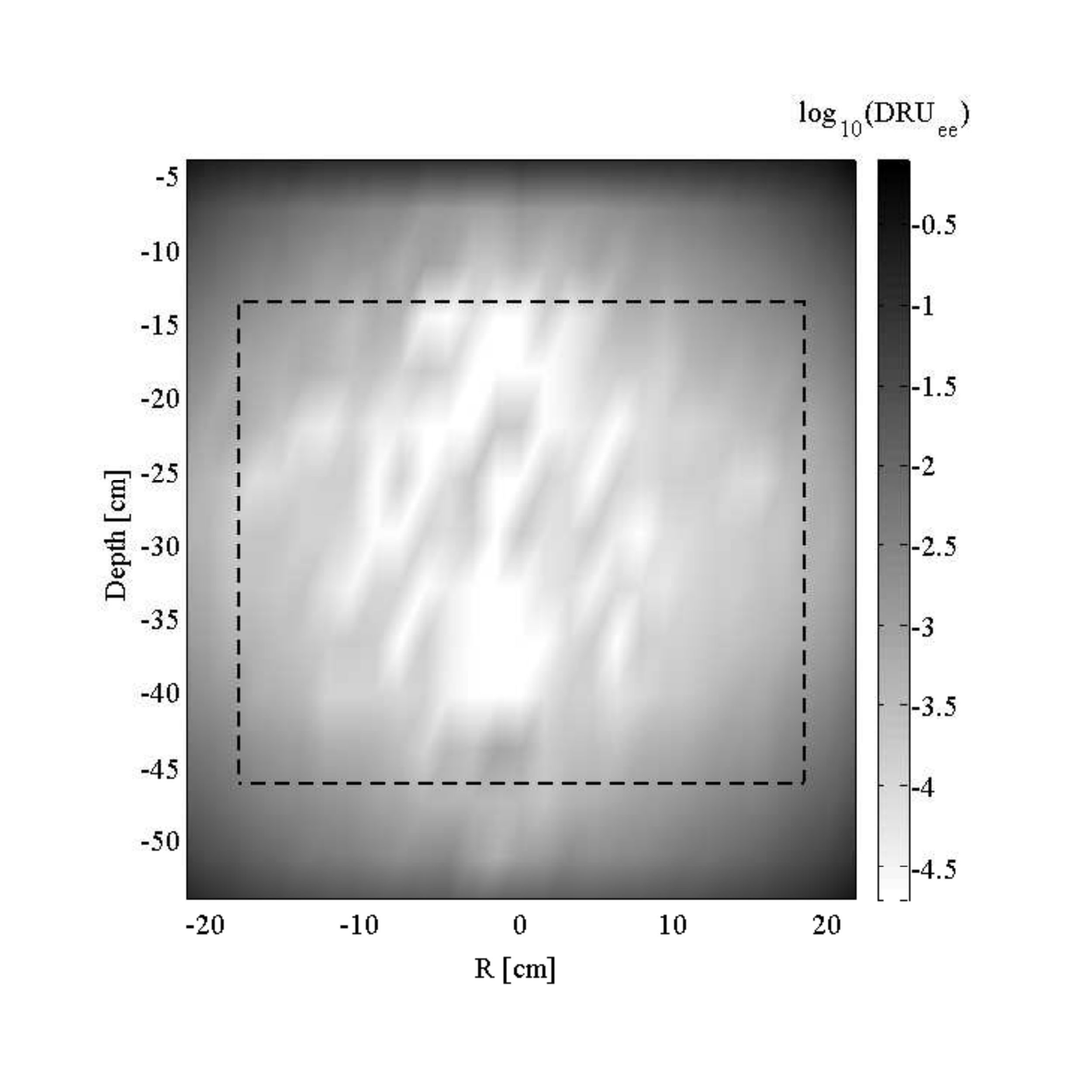} 
	\caption{Upper limit on the expected single-scatter ER event rate for energies between 5 and 25~keV$_{\text{ee}}$,  arising from the radioactivity of the R8778 PMTs, in terms of log(DRU) where 1 DRU~=~1 event/kg/keV$_{\text{ee}}$/day.  The dashed rectangle outlines the fiducial volume of the detector.\label{luxPMTbkgr-Fig}}
\end{center}
\end{figure}

The combined goal for $\gamma$ and $\beta$ rates in the LUX fiducial volume is $<$8x10$^{-4}$ events/kg/day/(keV electron equivalent or keV$_{\text{ee}}$) at threshold energy.  The 3D-imaging capability of the detector is a powerful tool to eliminate the major backgrounds in the outer portions of the detector, and can also be used to eliminate resolvable multiple-scattering events.  The effectiveness of these rejection methods depends strongly on the specific detector design.  Extensive Monte Carlo simulations have been performed for the LUX detector, and show that the necessary suppression can readily be achieved \cite{Akerib2012LuxSim}.  Fig.~\ref{luxPMTbkgr-Fig} shows the upper limit of the single-scatter ER event rate associated with the radioactivity of the PMTs, which dominates the LUX background rate.  The expected background rate due to the PMTs in the fiducial volume is $<$~10$^{-3}$ events/kg/keV$_{\text{ee}}$/day.  More details on background rates in LUX are provided in Sec.~\ref{sec:bg}.
	
\begin{figure}[]
\begin{center}
	\includegraphics[width=\columnwidth]{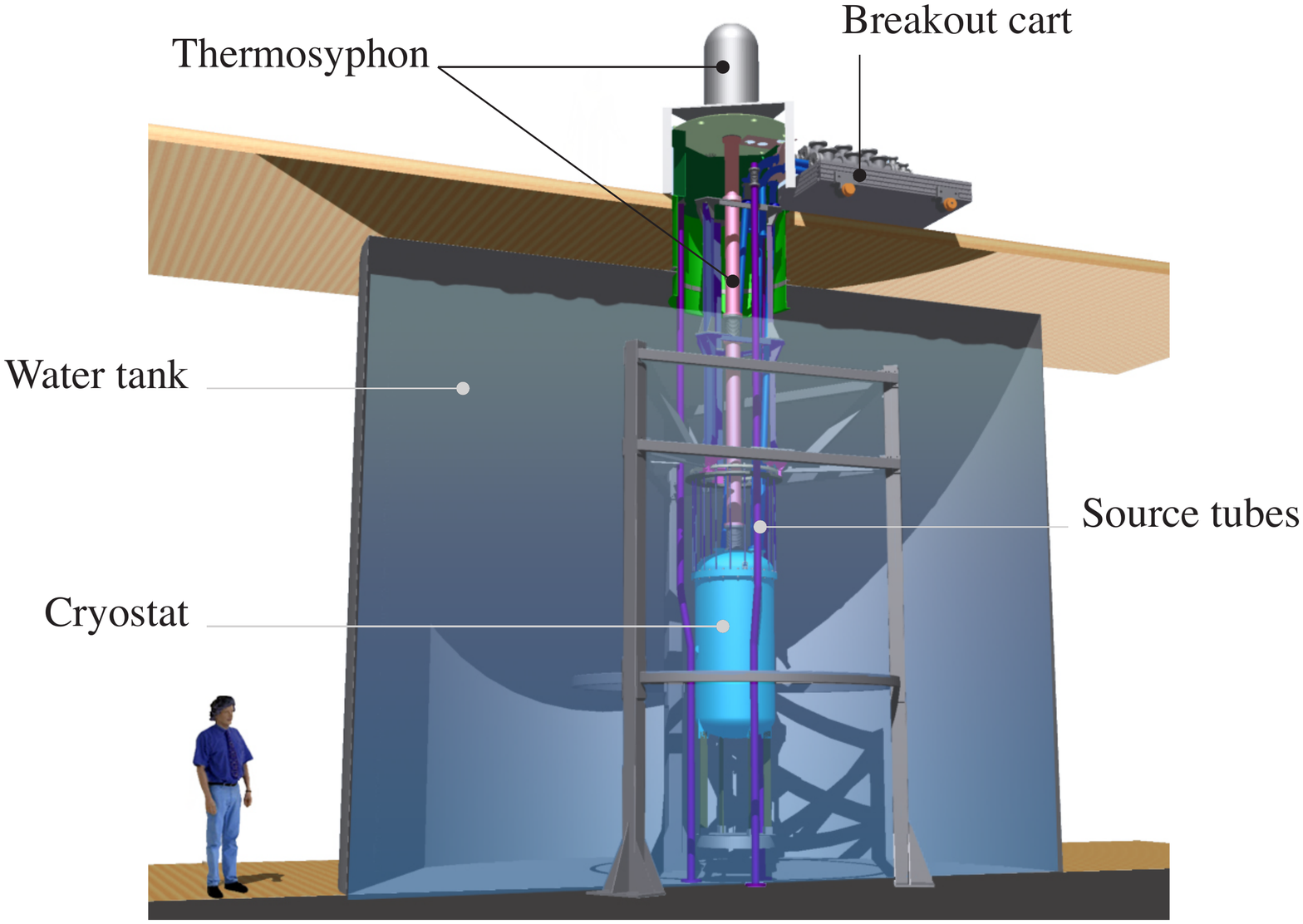} 
	\caption{Overview of the LUX detector system installed in the Davis Cavern.  Shown are the water tank and the central cryostat.  The PMTs of the muon-veto system are not shown.\label{lux-ug-Fig}}
\end{center}
\end{figure}

A schematic of the LUX detector system is shown in Fig.~\ref{lux-ug-Fig}.  The outer portion of the detector is a large water shield, discussed in more detail in Sec.~\ref{sec:watershield}, instrumented with PMTs, and used as a Cherenkov veto for muons.  The rate of single-scatter NR events in the LXe fiducial volume from muon-generated neutrons in the water or surrounding materials is $\sim$1~event per 3 years, before applying the veto.  Nevertheless, we want to ensure that any event associated with muons can be eliminated and also demonstrate that any WIMP candidate is not correlated with muons.

\begin{figure}[]
\centering
\includegraphics[width=\columnwidth]{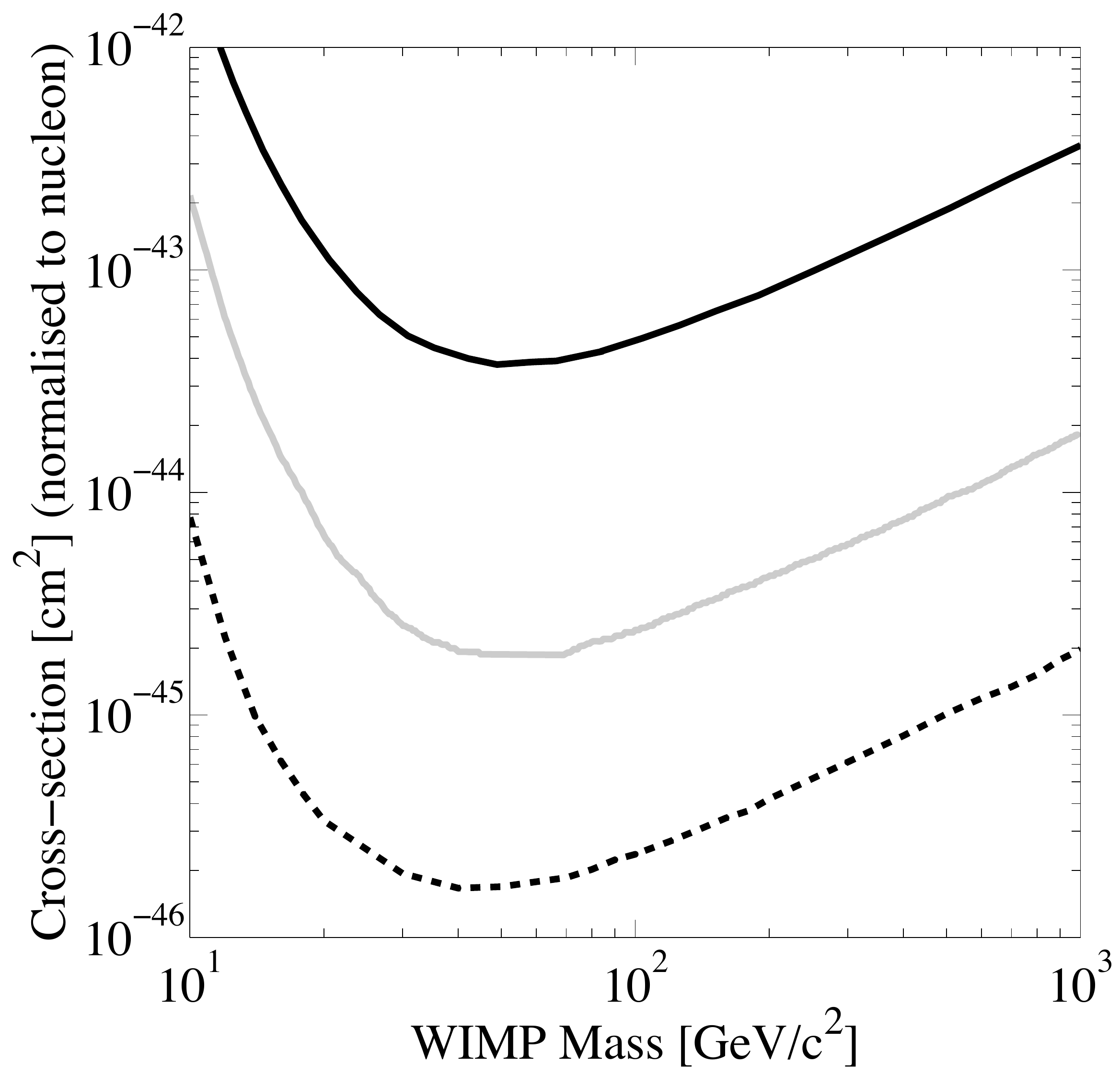}
\caption{LUX projected WIMP spin-independent sensitivity at the 90\% confidence level (dashed, black) after an exposure of 10,000 kg-days. Also shown are current limits from ZEPLIN-III (solid black, Ref.~\cite{AKimov2012}) and XENON100 (solid grey, Ref.~\cite{Aprile2012}).  Plot generated using the software of Ref.~\cite{DMTools}.}
\label{limit}
\end{figure}

The projected sensitivity of LUX,  after running for 100 days in dark matter search mode (10,000 kg-days of exposure), is shown by the dashed curve in Fig.~\ref{limit}. Also shown are limits from ZEPLIN-III (solid black, Ref.~\cite{AKimov2012}) and XENON100 (solid grey, Ref.~\cite{Aprile2012}).

\section{Detector Internals}
\label{sec:internals}

\subsection{Cryostat}
\begin{figure}[]
\begin{center}
	\includegraphics[width=\columnwidth]{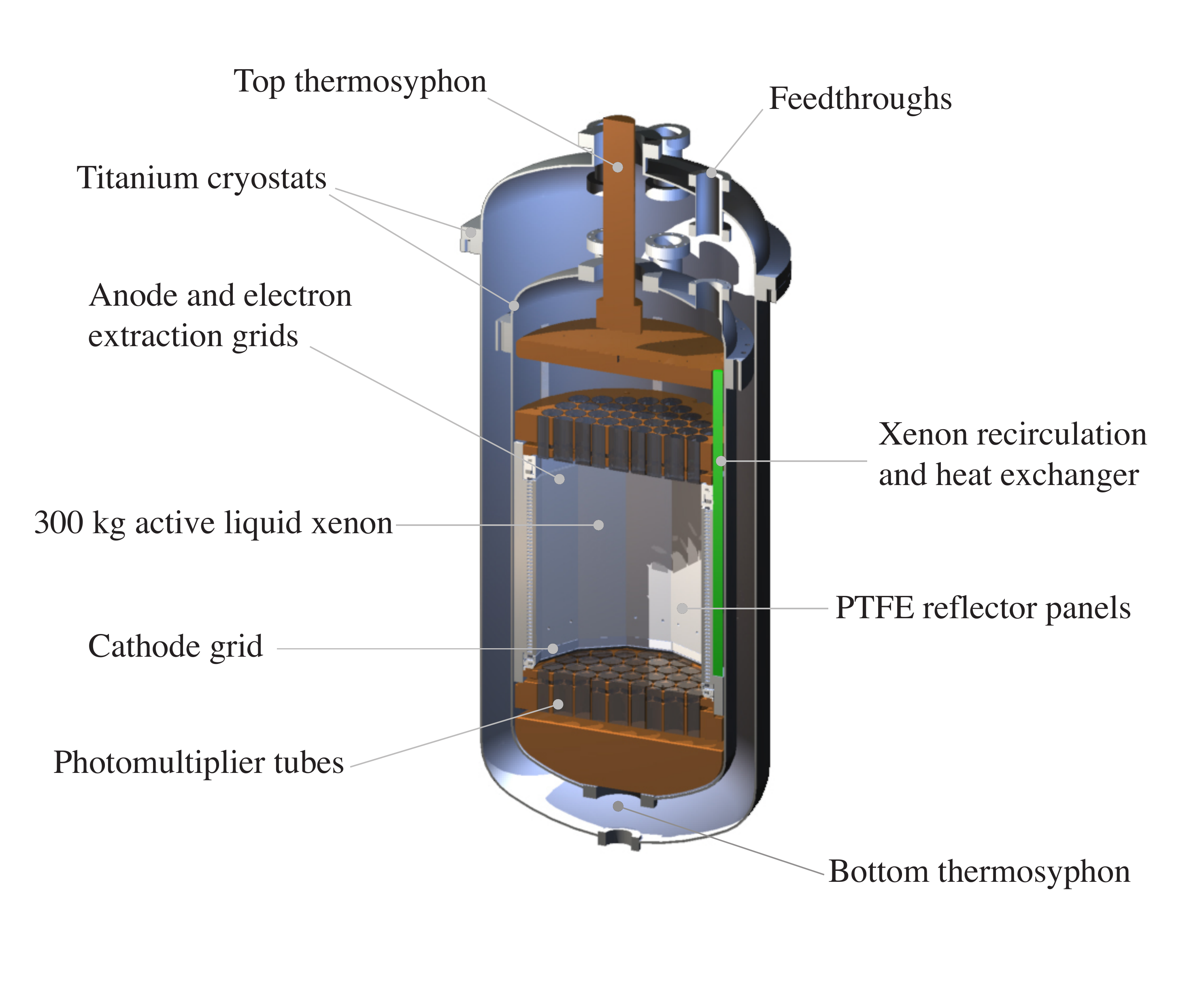} 
	\caption{Cross-sectional view of the LUX cryostats.  The vertical distance between the inner faces of the top and bottom PMT arrays is 61.6~cm.\label{lux-half-Fig}}
\end{center}
\end{figure}

The LUX detector cryostat vessels, schematically shown in Fig.~\ref{lux-half-Fig}, are made of 0.223" thick grade CP1 titanium sheets \cite{Akerib2012Ti}, machined and welded by Ability Engineering.  The outer vessel holds a vacuum to insulate the inner vessel while the inner vessel contains the detector internals and the liquid xenon.  The inner vessel is a 39.75-inch tall, 24.25-inch diameter cylinder with a dome welded to the bottom and a 27.75-inch diameter flange welded to the upper rim.  It hangs from the upper dome of the outer vessel and is mechanically attached and thermally isolated via plastic in three hangers.  The instrumentation wiring and circulation plumbing lines penetrate through the outer vessel into the inner vessel via commercially available stainless steel flexible couplings to compensate for thermal contraction of the plastic in the hangers.  The couplings are formed of thin-walled stainless steel and have low thermal conductivity.  The flanges of both vessels are designed to use Helicoflex gaskets.

\subsection{Cryogenics}

\begin{figure}[]
\begin{center}
	\includegraphics[width=\columnwidth]{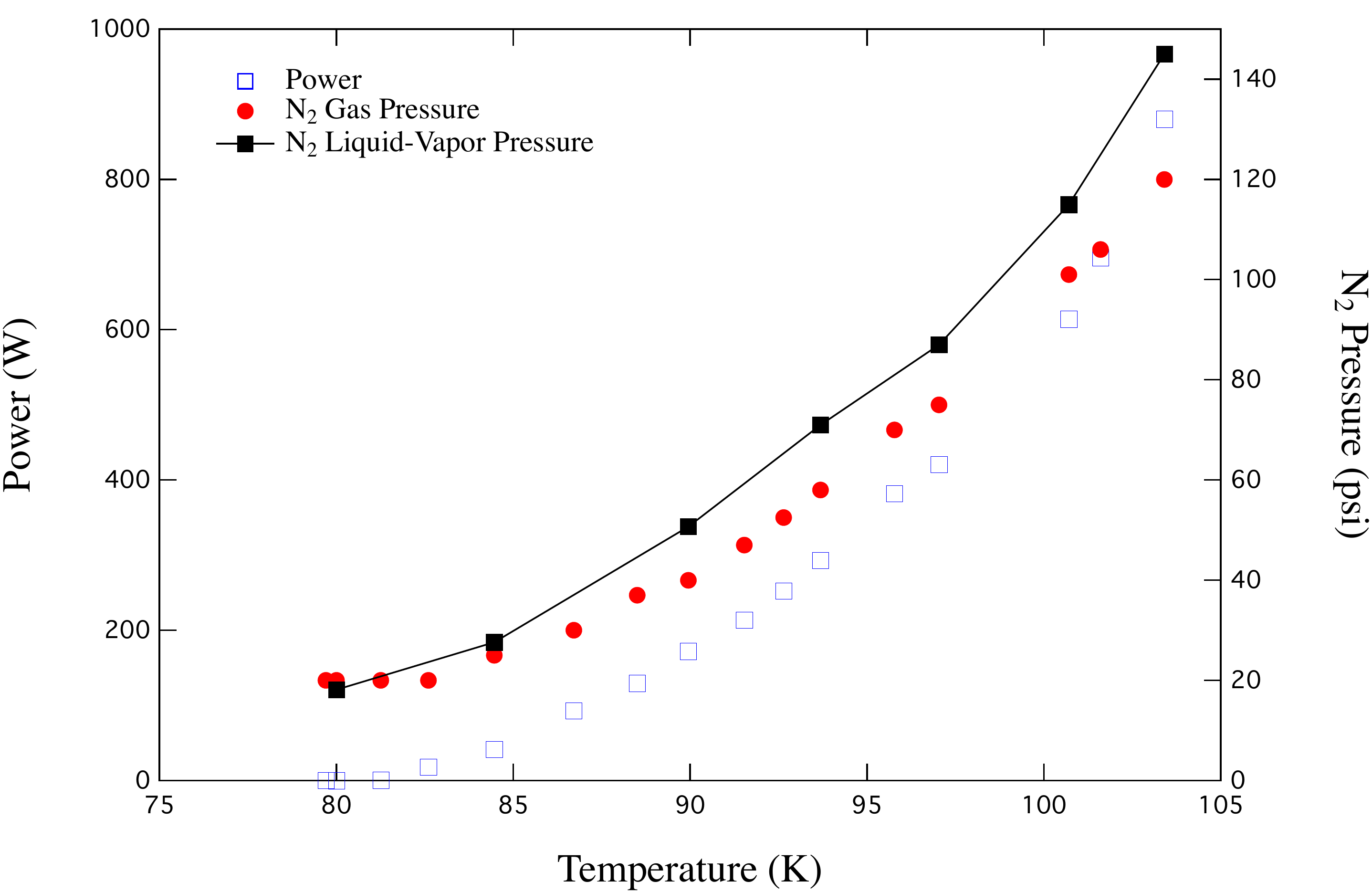} 
	\caption{Thermosyphon temperature as a function of power applied to the cold head.  Shown are the temperature of the cold head (open squares), the gas pressure  inside thermosyphon (solid circles), and the vapor-liquid equilibrium pressure for nitrogen  at the temperature of the cold head (solid squares).  The error bars are similar to the size of the data points.  Data taken from Ref.~\cite{Bolozdynya2009}.\label{TSeffic}}
\end{center}
\end{figure}

To efficiently and economically cool the LUX detector we use a unique cryogenic system based on thermosyphon technology \cite{Bolozdynya2009}.  Each thermosyphon consists of a sealed tube, partially filled with a variable amount of gaseous nitrogen (N$_2$), and comprised of three regions: at the top, a condenser which is immersed in a bath of liquid nitrogen (LN); at the bottom, an evaporator which is attached to the detector; and connecting these two active sections, a passive length made of stainless steel.  The thermosyphons are oriented vertically since they work with gravity, and are closed and pressurized with N$_2$.  The N$_2$ condenses inside the condenser and trickles down the stainless steel lines to the copper cold heads (evaporators) that are securely fastened to various points on the detector's inner can. The nitrogen evaporates, removing heat from the detector. It then rises back up the lines towards the condenser.  Figure~\ref{TSeffic} shows that the pressure in the thermosyphons is slightly below the equilibrium vapor pressure corresponding to the temperature of the evaporator.   This condition allows for the highly efficient operation of the thermosyphons.  Measurements of the thermosyphon thermal conductivity put it at $\sim$55~kW/K/m, much higher than copper, and comparable to carbon nanotubes at low temperatures (see Ref.~\cite{Bolozdynya2009} and references therein). This technology has demonstrated excellent efficiency in cooling, as shown in Fig.~\ref{TSeffic}.  

Four thermosyphon cold heads are deployed in the LUX detector.   One large capacity cold head is placed at the top of the inner vessel and thermally coupled to the top copper thermal and radiation shield.  A smaller capacity cold head is placed at the bottom of the inner vessel and thermally coupled to the bottom copper thermal and radiation shield.  These two cold heads are used to cool the detector internals slowly and uniformly from room temperature to 175 K.  The other two cold heads are attached to the copper thermal shield surrounding the inner vessel and are used to maintain a thermal gradient along the vertical dimension of the detector.  The cold heads can convey $\sim$200-400~W of cooling power.  Each head is instrumented with a 50-W heater and a thermometer as part of a Proportional-Integral-Derivative (PID) loop for fine temperature control.  The top and bottom cold heads are additionally instrumented with a 750-W heater and a PID control thermometer to assist with warming the detector internals and LXe recovery.

\subsection{Grids, fields, and light reflecting cage}
\label{sec:grids}

\begin{figure}[]
\centering
\includegraphics[width=\columnwidth]{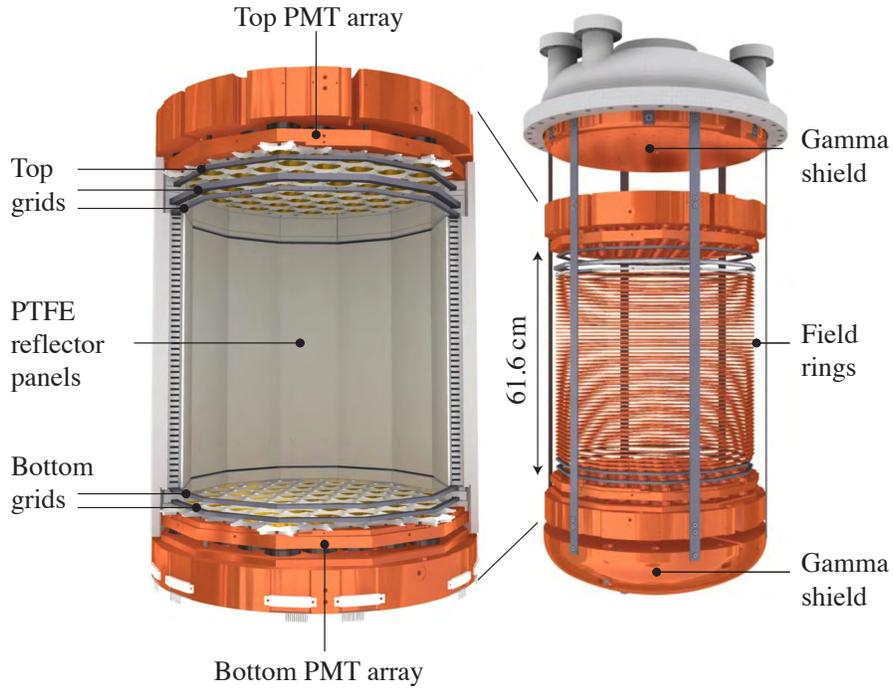}
\caption{Rendering of the LUX TPC, supported from the top flange of the inner cryostat.}
\label{Internals}
\end{figure}

The LUX Time-Projection Chamber (TPC) is a dodecagonal structure enclosing an active region with approximately 300 kg of liquid xenon. The active region is viewed from above and below by arrays of 61 PMTs, as illustrated in Fig.~\ref{Internals}. Monte Carlo optimization of background rejection and fiducial volume resulted in a design with a drift distance of 49~cm, a diameter of 50~cm, and a buffer distance of 5~cm between the cathode and the bottom PMT array. The inner walls of the TPC consist of twelve polytetrafluoroethylene (PTFE) reflector panels that cover forty-eight copper field rings, supported by Ultra High Molecular Weight Polyethylene (UHMW) panels. All PTFE components are made from ultrahigh purity grade materials and all copper components are C101 OFHC grade. The field cage includes five grids, supported by PTFE structures, that maximize light collection and minimize the leakage of scintillation light from xenon outside the TPC into the viewing region.  The entire structure is supported from the top flange of the inner vessel, as shown in Fig.~\ref{Internals}. A 5~cm thick copper disk with a diameter of 55~cm is mounted directly on to the flange. It connects to the large capacity thermosyphon through a cold finger and is used for temperature control and heat sinking of cables, and serves as a $\gamma$ shield.  All other components are supported from six titanium straps (alloy CP1) attached to this shield. The top PMT array is supported 15~cm below the shield in a copper structure and PTFE reflectors cover the regions between the faces of the PMTs.  The polyethylene panels hang from the PMT support and are attached to the bottom PMT support with slotted holes to allow for thermal contraction when cooled. All components were designed to remain stress free during baking at temperatures up to 40$^{\circ}$C and cooling to liquid xenon temperatures. The bottom PMT support is a copper structure, similar to the top PMT support, with PTFE reflectors covering the gaps between the PMT faces as well. Below the bottom PMT support, a 15 cm thick copper structure fills the dome shape of the inner vessel to displace inactive xenon and to provide additional $\gamma$ shielding. This copper structure is also connected to a thermosyphon to provide control of the temperature gradient in the detector and to adjust the temperature of the returning xenon in the circulation system.   

The lowest grid is located 2~cm above the bottom PMTs to shield the photocathodes of the PMTs from the high voltage on the cathode grid. It employs 206~micron diameter ultra-finished stainless steel wires spaced with a pitch of 1~cm and supported by a thin stainless steel ring. Except for the screws, all stainless steel used in grid construction is alloy 316. This grid has an open area of 98\% and its voltage is adjusted to ÓzeroÓ the electric field at the photocathodes of the PMTs. The gauge and pitch of the wires were selected to achieve the required shielding while maintaining the mechanical strength to allow a voltage of up to 100~kV on the cathode. The cathode grid is located 4~cm above the bottom PMT shield and is identical in construction, except that it has a pitch of 5~mm and a 96\% open area. These two grids were tested at the design spacing with 100~kV between them to verify mechanical stability. The liquid xenon surface is nominally 49.5~cm above the cathode grid. Sitting $\sim$5~mm below the surface of the liquid is the ÒgateÓ grid that is used in conjunction with the anode to generate the 5~kV/cm extraction field just below the liquid surface and the 10~kV/cm electro-luminescence field in the gas above the liquid.  The gate grid has a pitch of 5~mm and uses 50 micron stainless steel wires, resulting in a 99\% open area. The pitch was optimized to result in a variation in the extraction field across the surface of the liquid of less than 1\%. The anode sits a distance of 1~cm above the gate grid and is constructed using a stainless steel mesh with 30 micron wires, spaced by 0.5~mm. The open area of the anode grid is 88\% and the structure supporting the mesh allows tensioning of the mesh after it is mounted on to the frame. After tensioning, the deflection of the mesh was measured to be 130 microns at the design electroluminescence field. The top grid is located 4~cm above the anode and 2~cm below the top PMT array. Its function and geometry are identical to the bottom PMT shield, except that it is strung with 50 micron wires, resulting in a 99\% open area. 

The copper rings used to shape the drift field have a thickness of 3.2~mm, a width of 12.7~mm, and a spacing of 1~cm. They are held in grooves in the polyethylene panels behind the PTFE reflector panels. The spacing and thickness of the rings, along with the location and voltage step to the cathode and gate, were chosen to provide as uniform a drift field as possible while shielding the region from the electric field generated by the cathode HV cable.  

A resistor chain between the gate and the cathode grids is used to fix the voltage of the field rings.  Adjacent field rings are connected by a parallel pair of 1~G$\Omega$ resistors.  A parallel pair of 0.875~G$\Omega$ resistors connects the top field ring to the gate grid and a parallel pair of 1.25~G$\Omega$ resistors connects the bottom field ring to the cathode grid.

\subsection{Cathode HV}
The LUX high-voltage system can deliver up to 100~kV to the cathode in order to create a 2~kV/cm drift field in the liquid. This is particularly challenging because high-purity noble gases will readily discharge. Rather than design a custom, low-radioactivity feedthrough for low-temperature operation, as was done in the ICARUS and WARP experiments, the LUX cathode high-voltage system places the feedthrough outside the water shield at room temperature.  The feedthrough forms a vacuum seal between commercial high-voltage cable (HVC100, made by Heinzinger Electronic) and a standard conflat flange.  One side of the feedthrough is surrounded by transformer oil (PMX-561, made by Xiameter), while the other side is in xenon gas and connects to the high-voltage cable, which passes through a xenon-gas-filled umbilical into the liquid xenon detector.  The cable is shielded until it is 25~cm deep into the liquid xenon, where the ground shielding is removed. The cable then continues another 30~cm down to the cathode, where the final connection is made with a spring-loaded pin, surrounded by liquid xenon and polyethylene. In order to avoid back-diffusion of outgassed impurities from the high-voltage cable into the detector, xenon gas is drawn away from the detector through the umbilical and returned to the xenon purification system at a controlled rate of 0.1 standard liters per minute (SLPM).

The system avoids discharges in the xenon by excluding gas from all high-field regions.  Within the transformer oil, a current limiting resistor is added in series to minimize the harmful consequences of a short. A second cable connects this resistor to the power supply.

\subsection{Internal circulation system}
\label{sec:internalCirculation}

The LUX circulation system was designed to allow for the continuous purification of the xenon target to maintain the required electron drift length necessary for full 3D imaging within the detector.  This purification is accomplished through the use of a SAES hot metal getter (see Sec.~\ref{sec:gas} and Ref.~\cite{SAESMonoTorr}).  As the SAES getter requires gaseous xenon, the xenon is continuously evaporated and recondensed in the circulation system.  LUX uses a series of heat exchangers to mitigate the heat load of this process.

\begin{figure}
\centering
\includegraphics[width=1\textwidth]{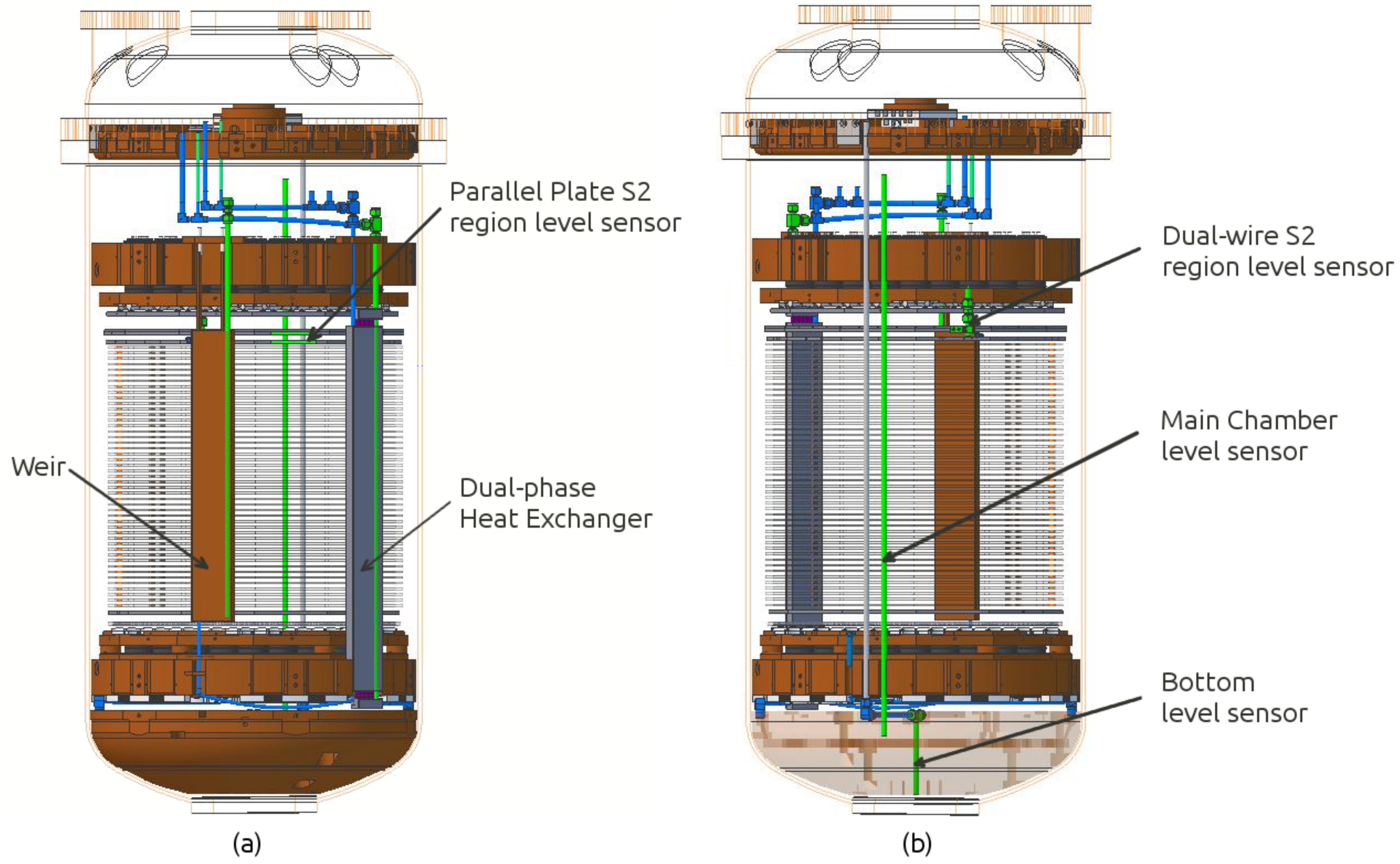}
\caption{Rendering of the LUX central cryostat with its reflector panels suppressed.  The circulation path and level sensors are shown. \textbf{(a)} Detector from 11:00 position.  Only one parallel-plate level sensor is shown. \textbf{(b)} Detector from 5:00 position.  The lowest copper section is rendered transparent for clarity.}
\label{circulationSchematic}
\end{figure}

The fluid path through these heat exchangers during circulation, shown in Fig.~\ref{circulationSchematic}, is as follows.  Starting from the top of the active region, liquid spills over the weir lip into the weir reservoir.  The weir reservoir is a 
buffer volume designed to handle any changes in effective liquid level due
to changes in temperature or circulation rate.  The level
of the liquid in the weir is monitored using a capacitive level
sensor.
From the weir, the liquid xenon flows into the evaporator side of a dual-phase heat exchanger and evaporates.  The gaseous xenon flows out of the detector through a single-phase concentric-tube heat exchanger, which warms the gas to nearly room temperature.  A diaphragm pump then pumps the xenon through the getter.  In order to avoid contamination of xenon by air, the inter-diaphragm space is pumped on. After purification, the gas flows back into the detector, first passing through the concentric tube heat exchanger, where it cools to near its condensation point.  Entering the condenser side of the dual-phase heat exchanger, the gas  condenses and drains down into the lower $\gamma$-ray shield.  Here, the condensed xenon is fed through a channel to ensure it is at the same temperature as the rest of the detector before it is reintroduced at the bottom of the active region.  

\begin{figure}
\centering
\includegraphics[width=.75\textwidth]{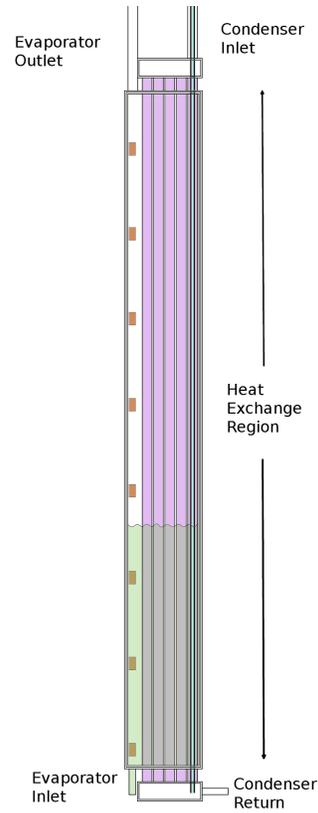}
\caption{LUX dual-phase heat exchanger.  A total of eight thermometers, equally spaced along the vertical dimension of the heat exchanger,  are shown as small grey rectangles on the left-hand side.  The five condenser tubes are shown to the right of the thermometers.  A single capacitance level sensor of the parallel wire type is shown on the right-hand side of the heat exchanger.}
\label{figHX2}
\end{figure}

The dual-phase heat exchanger, shown in Fig.~\ref{figHX2}, consists of a series of five tubes inside an encapsulating space.  This heat exchanger mitigates the heat load that is incurred due to gaseous xenon condensation.  Liquid xenon enters the evaporator side of the heat exchanger from the weir reservoir, is pumped on by the external circulation system, and evaporates.  This evaporation creates a cooling surface which is used to condense xenon gas being returned by the external circulation system.  This heat exchanger technology has been demonstrated to work at greater than 94\% efficiency, up to a rate of 350 kg/d \cite{AkeribLUX01}.

The dual-phase heat exchanger is instrumented with a variety of sensors for diagnostic and control functions.  As the circulation rate changes, the overall liquid height in the evaporator changes.  Thermometers are used in addition to a differential pressure measurement to gauge the liquid height because evaporation leads to the creation of froth which renders capacitance level meters unreliable.  Inside the condenser, a single parallel-wire capacitance based level sensor is used to gauge the liquid level.

\subsection{Internal instrumentation}
\label{sec:internalInstrumentation}

\subsubsection{Thermometers}
\label{sec:thermometers}

The temperature sensors used in LUX consist of thin film platinum resistors ($\unit[100]{\Omega}$) deposited onto a ceramic substrate from Omega (part number F2020-100-1/3B-100, Class AA tolerance \cite{OmegaWeb}). The precision of these thermometers is quoted as 100~mK at 0$^{\circ}$C.
The majority of these thermometers are monitored with Advantech ADAM modules (model \# 6015) that use a three-wire measurement technique \cite{AdvantechWeb}.
The specified intra-channel accuracy of any given module is \unit[400]{mK}.   

The majority of thermometers are mounted on Cirlex$^{\tiny{\textregistered}}$ boards \cite{CirlexWeb} such that the ceramic substrate of the thermometer is pressed against the surface being monitored using a stainless steel 6-32 screw and a lock-washer. 
The flexibility of the lock-washer ensures that the ceramic substrate of the thermometer is always in good thermal contact with the monitored surface. 
A total of 40 thermometers are installed in he xenon space and 23 are mounted in the vacuum space.

Prior to installation, each thermometer was cycled 10 to 15 times between 300 K and 77 K. 
The calibration procedure consisted of mounting 14 thermometers on an aluminum block, covered with a heat shield, and allowing the system to equilibrate while sitting in various temperature baths, such as liquid nitrogen, dry ice and ethyl alcohol, and  room temperature. 
One thermometer was chosen as the standard and temperature differences from the standard thermometer were recorded. 
At each temperature, data were collected for one minute. Typically, each thermometer was cycled five times through each temperature bath.

Surprisingly, it was observed that the temperature differences from the standard thermometer had a quadratic dependence upon temperature, a dependence upon the particular triplet of wires used in the calibration process, and a dependence on the particular ADAM module used to make the measurement. 
Furthermore, we observed a high failure rate (40\%) of thermometers during the calibration process.  Several thermometers reported irreproducible results and/or differences with the standard thermometer of several kelvin.   It was determined that the failure rates was significantly higher for thermometers with a non-standard thermometer mount .  A redesign of this mount resulted in a significant reduction of the failure rate.

Experiments were carried out to determine the effect that the wiring and the ADAM module have on the temperature measurements. 
The observed discrepancy between temperature measurements made with two different ADAM modules is $\unit[60\pm 9]{mK}$. The corresponding discrepancy due to the wiring is $\unit[147\pm47]{mK}$.   
The total RMS value of the temperature offsets of thermometers installed in the xenon space was determined to be $\unit[780\pm 270]{mK}$. After applying the proper calibration constants and corrections, the accuracy of the temperature measurements in LUX is determined to be 170~mK.

\subsubsection{Level sensors}
\label{sec:levelSensors}

Two varieties of liquid level sensor are deployed in LUX, as shown in Fig.~\ref{circulationSchematic}.  In the circulation plumbing, where unimpeded fluid flow is important, parallel-wire level sensors are used.  Level sensors of this type are used to measure the liquid level in the main chamber, the weir, the condenser of the dual-phase heat exchanger, and the liquid return line that reintroduces liquid to the active region after circulation.  These sensors consist of two parallel wires, mounted in compression-type connections, and designed to fit into standard Swagelok$^{\tiny{\textregistered}}$ fittings.  Measuring the capacitance of each wire pair allows the length of the submerged portion of the wires to be determined.

A different type of level sensor is used to measure the liquid xenon level in relation to the electric field grids.  To achieve the level of accuracy required, three level sensors of a parallel-plate design were incorporated into the structure surrounding the active region between the gate and the anode grids.  These sensors, spaced 120 degrees apart, surround the electron extraction region and provide a method to determine if the liquid surface is level with respect to the grids, thus ensuring a uniform extraction field.

\subsection{Slow control}
\label{sec:controls}
To monitor detector health and control operations, a unified slow control
system has been implemented that covers all of the LUX instrumentation.
The detector temperature is monitored in 61 separate places.  Thermometers are also placed on external 
equipment.  The liquid xenon level in the detector is monitored in nine
different locations and the xenon gas system is fully instrumented with
flow controllers and pressure gauges.
The PMT and grid high-voltage supplies, residual gas
analyzers, and liquid nitrogen systems are also monitored and
controlled through the same interface.  All data is stored permanently
in a master database which is mirrored both on-site and off-site.  The
primary user interface is through a web browser, though the flexibility
of the database structure means that many other interfaces are possible.
Both ROOT and MATLAB interfaces to the slow control database have been implemented for specialized
analysis tasks.

A set of alarm and watchdog systems alert people, both on and off
site, with lights, sirens, emails, and text messages, depending on
the severity of the situation.  Each sensor can be associated with any
combination of high or low (value or rate) alarm trigger points, and
alerts are routed to those responsible for the system that triggered the alarm.

In total more than two hundred sensors are monitored and dozens of
instruments are controlled by the slow control system.  The system was
originally designed for the DEAP/CLEAN collaboration and has been
deployed in numerous other experiments, including the LUX 0.1
prototype detector \cite{AkeribLUX01}.

\section{Detector Calibration System}
\label{sec:calibrations}
Calibrations of the LUX detector are used to determine the predicted WIMP signal and the detector's ability to distinguish between nuclear and electron recoils. The calibrations are performed using sealed $\gamma$-ray and neutron sources as well as internal $^{83m}$Kr \cite{Kasten2009,Manalaysay2010} and Tritium beta sources. For electron-recoil calibrations we employ $\gamma$ rays covering a broad energy range from about one hundred keV to a few MeV, as well as beta electrons in a range from a few keV to a few tens of keV from internal sources. Low-energy $\gamma$-ray sources (e.g. $^{57}$Co) and $^{83m}$Kr are used to define the electron-equivalent energy scale (keV$_{\text{ee}}$) and for xenon purity monitoring. High-energy $\gamma$-ray sources, such as $^{137}$Cs, $^{22}$Na, $^{208}$Tl, and Tritium, are used to determine the boundaries of the electron-recoil energy band in the S2/S1 vs S1 plane. The nuclear-recoil energy band and the nuclear recoil scintillation yield are studied using neutron sources. In order to reduce the attenuation of the $\gamma$-ray and neutron flux by the water, the radioactive sources are placed inside the LUX water tank, in close proximity to the outer cryostat. The $\gamma$-rays are collimated to reduce the number of interactions in the layer of LXe outside the region of interest. 

\begin{figure}[]
\centering
\scalebox{0.5}{\includegraphics[width=\columnwidth]{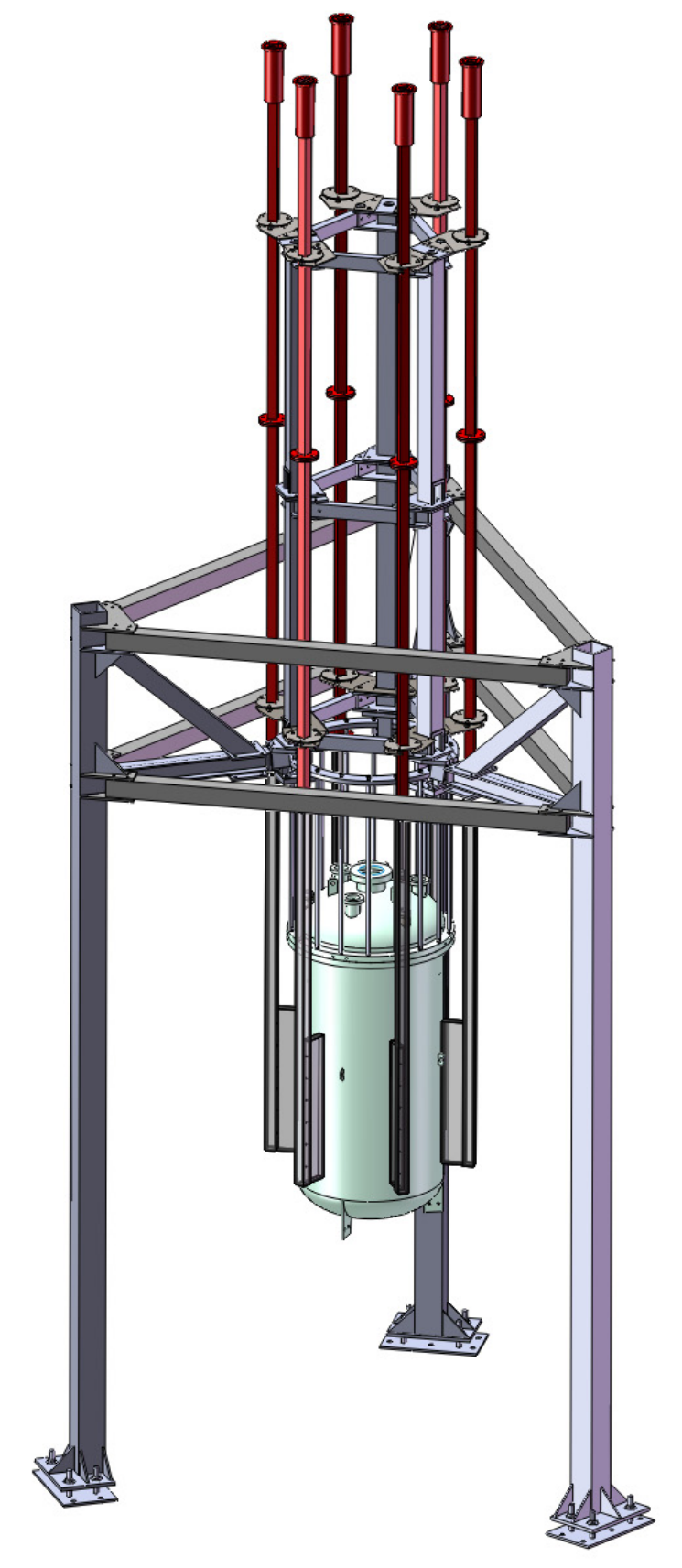}}
\caption{Rendering of the six source tubes surrounding the central LUX cryostat.  The top two sections of each source tube are made of steel; the bottom section is made of clear acrylic.  Clear acrylic water displacers are installed in front of the active xenon volume to limit source attenuation.}
\label{sourceTower}
\end{figure}

\begin{figure}[]
\centering
\includegraphics[width=\columnwidth]{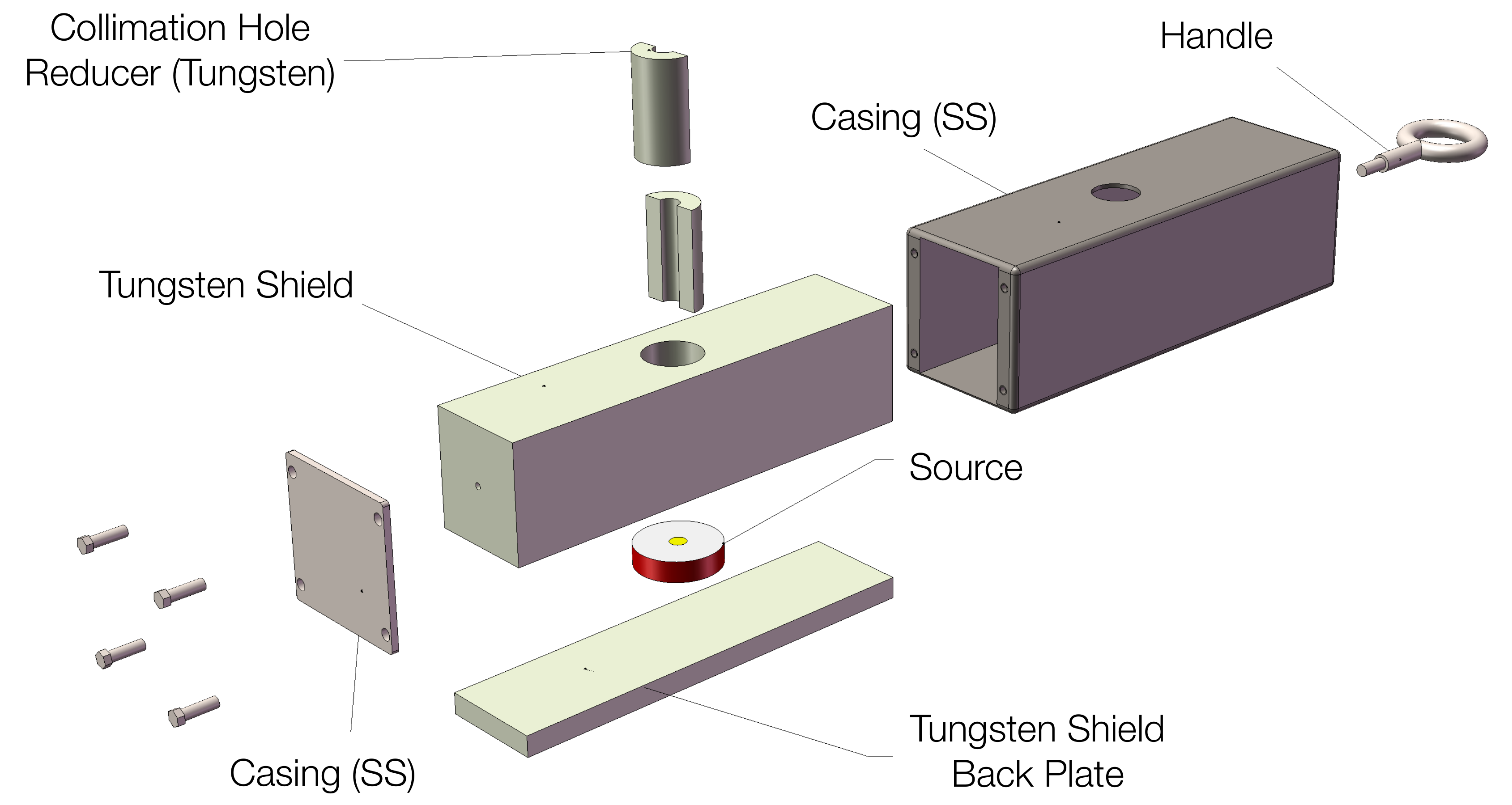}
\caption{Gamma-ray source assembly used for detector calibrations.}
\label{sourceHolder}
\end{figure}

The calibration system for sealed sources includes six dry source-tubes, shown in Fig.~\ref{sourceTower}, and collimator assemblies for $\gamma$-ray sources, shown in Fig.~\ref{sourceHolder}. Equally spaced at an appropriate radius from the center of the cryostat, the source tubes drop down from the top of the tank to the bottom of the cryostat. They are attached to the cryostat stand using stainless steel mounting brackets. Each tube has two mounting flanges which rest on the brackets using flexible mounts, permitting adjustment of the orientation of the tubes. The inside of the source tubes is kept dry.  The amount of air in the tubes produces a negligible background in the detector.  The source tubes have a square cross-section to fix the orientation of source holders as they move inside the source tubes. The collimation holes in the source holders vary from 0.5~mm to 17~mm. Large collimation holes are used for measurements of the electron-recoil energy and the electron-equivalent energy scale while narrow collimator holes are used for position-resolution measurements. A tungsten backing is installed behind the source in order to minimize the number of photons that enter the active volume of the detector after backscattering in the source assembly. To study the detector response as a function of height, the sources can be moved up and down along the tube, using a mechanism with reels and pulleys. Three collimator assemblies for each collimation hole are available, allowing for the simultaneous irradiation of the active volume of the detector from three different sides.

Internal sources, such as $^{83m}$Kr and tritiated methane, will be introduced by doping the gaseous xenon with the radioactive isotopes. In the case of $^{83m}$Kr, the isotopes will decay away within 1.86 hours. Tritiated methane will be removed from the xenon using the getter. We have determined that 99.9\% of the tritiated methane is removed in a single pass throughout the purification system. Details about calibration techniques using internal sources can be found in Refs.~\cite{Kasten2009,Manalaysay2010}.

\section{Purification and Gas Handling}
\label{sec:gas}
The LUX purification system removes electronegative and molecular
impurities that affect charge and light collection using gas-phase
recirculation through a commercial heated getter (SAES MonoTorr \cite{SAESMonoTorr}),
developed for the semiconductor industry.  This is the same approach as
has been used in ZEPLIN-II, XENON10, and LUX 0.1 \cite{AkeribLUX01} where absorption
lengths of several meters have been achieved. This is sufficient as a
baseline for the 49-cm depth of LUX. The xenon evaporates from the
detector and is circulated through the getter using a
double-diaphragm pump.  The xenon flow rates and pressures are monitored
by the slow control system.

The LUX purification system is capable of processing 50 SLPM (420~kg/day).  In order to
deal with the associated 400~W heat load, LUX employs internal
heat exchangers, described in Sec.~\ref{sec:internalCirculation}, that use the cooling provided by xenon evaporation to
condense the returning gas. These heat exchangers, developed for LUX, result in a demonstrated 98\% reduction in required cooling
power \cite{AkeribRun2}.  Increased xenon purification rate is accomplished with
commercially available pumps, plumbing, and getters that allow higher
xenon gas throughput.  Xenon storage and recovery is handled
using a combination of a liquid nitrogen encased pressure vessel and
a balloon to allow emergency recovery of the xenon.

During operation of LUX on the surface, a circulation rate of 300~kg/day was achieved, limited by the sub-optimal circulation path used during surface operation \cite{AkeribRun2}.  The measured heat load was less than 5~W.

The xenon handling system is equipped with an integrated xenon gas
purity assay system. A 0.5 liter sample of xenon gas can be collected
from four critical locations in the system: the input to the getter,
the output from the getter, the detector return line, and a conduit
purge line. The gas samples are collected via plumbed-in stainless
steel lines which connect the xenon system to the assay system. The
impurity content of each sample is evaluated using the cold trap mass
spectrometry method which is described in Refs.~\cite{Leonard2010,Dobi2011}. This
technique is sensitive to oxygen and nitrogen at a level of better
than one part-per-billion, and can detect krypton at the
part-per-trillion level. The mass spectrometer is periodically
calibrated by collecting a sample of xenon gas from a specially
prepared cylinder which contains xenon with a known impurity
content. The assay system is used to monitor the effectiveness and
performance of the getter, to monitor the purification process, and to
confirm that the xenon gas system is leak tight.

\section{PMTs and Light Collection}
\label{sec:pmts}

LUX uses two arrays of 61 Hamamatsu R8778 PMTs \cite{Akerib2012PMT}. One array is placed above the liquid surface and is used primarily for x-y position reconstruction from the S2 signal pattern. Using likelihood pattern recognition techniques, developed for XENON10 \cite{Aprile2011} and ZEPLIN \cite{Solovov2012}, a $\sim$1~cm position reconstruction accuracy (20\% of the PMT diameter) is expected at low energy. The second array is placed in the liquid, below the cathode grid. This array gets most of the S1 scintillation signal, due to internal reflection at the liquid surface.

The R8778 is a 12-stage, 5.7~cm round PMT, developed by Hamamatsu and the XMASS collaboration specifically for operation in LXe. The photocathode is sensitive to 175 nm LXe scintillation, has a typical quantum efficiency (QE) of 33\% at that wavelength, can operate at 170 K, and can withstand up to 5 atm of pressure. Other relevant R8778 characteristics are listed in Table \ref{PMTspecs}.

Every PMT was biased and single photoelectron (sphe) pulses were recorded to confirm that they were all operational. The sphe pulse shape (rise time, fall time, FWHM, height) and sphe spectrum (gain, peak-to-valley ratio, sphe resolution) were evaluated for 25 PMTs at room temperature. The anode output linearity was  measured for one PMT, and the measurement agreed with the 13 mA at $\pm$2\% deviation specified by Hamamatsu. An extensive ion-feedback afterpulsing (AP) characterization study was performed for 90 PMTs, and the residual gas populations were identified for each of the 5 major peaks in the AP spectrum using each ion's mass-to-charge timing dependance. This allows us to track the PMT health throughout the experiment (see section \ref{PMTHealthMonitoring}). To ensure that the PMTs can operate in LXe, a 12 PMT (10\%) sample was tested in a LXe chamber. An AP measurement was performed before and during LXe immersion, and the Xe$^{+}$ peak was monitored to ensure that the PMT vacuum was not compromised during cooldown. None of the PMTs showed any sign of vacuum deterioration during LXe testing.

During the LUX 0.1 program \cite{AkeribLUX01}, four R8778 PMTs (three above the liquid surface, one below the cathode grid) were used to collect the scintillation light in the active LXe region for a period of 2 years. During that time, LED, $^{57}$Co, $^{133}$Ba, and $^{252}$Cf calibration measurements were successfully performed with them. The energy calibrations with $^{57}$Co and $^{133}$Ba allowed us to monitor the purity of the LXe by measuring the electron lifetime.

\begin{table}
\begin{centering}
\begin{tabular}
{|   l   |   c   | }
\hline			
  Minimum Photocathode Effective Area & 15.9~cm$^{2}$\\
  Typical voltage for a $2\times$10$^6$ gain$^\star$$^\dagger$ & 1100 V \\
  Maximum voltage$^\dagger$ & 1750 V \\
  Single photoelectron (sphe) rise time$^\star$ & 3.5 ns \\
  Sphe FWHM$^\star$ & 6.8 ns \\
  Sphe fall time$^\star$ & 10.8 ns \\
  Typical sphe spectrum peak-to-valley ratio$^\star$ & $\sim$4 \\
  Typical sphe peak resolution$^\star$ ($\sigma/\mu$) & 37\% \\
  Typical quantum efficiency (QE) at 175 nm$^\dagger$ & 33\% \\
  Standard deviation of the QE at 175 nm$^\dagger$ & 2.3\% \\
  Collection efficiency$^\star$ & $\sim$90\% \\
  Anode linearity at $\pm$2\% deviation$^\star$$^\dagger$& 13 mA \\
  Afterpulsing ratio at 1300 V in 5 $\mu s$$^\star$ & $<$5\% \\
\hline
\end{tabular}
\\ $^\star$Measured by LUX \\ $^\dagger$Measured by Hamamatsu
\par\end{centering}
\caption{Characterestics of the LUX R8778 PMTs. }
\label{PMTspecs}
\end{table}

\begin{figure}[]
\centering
\includegraphics[width=\columnwidth]{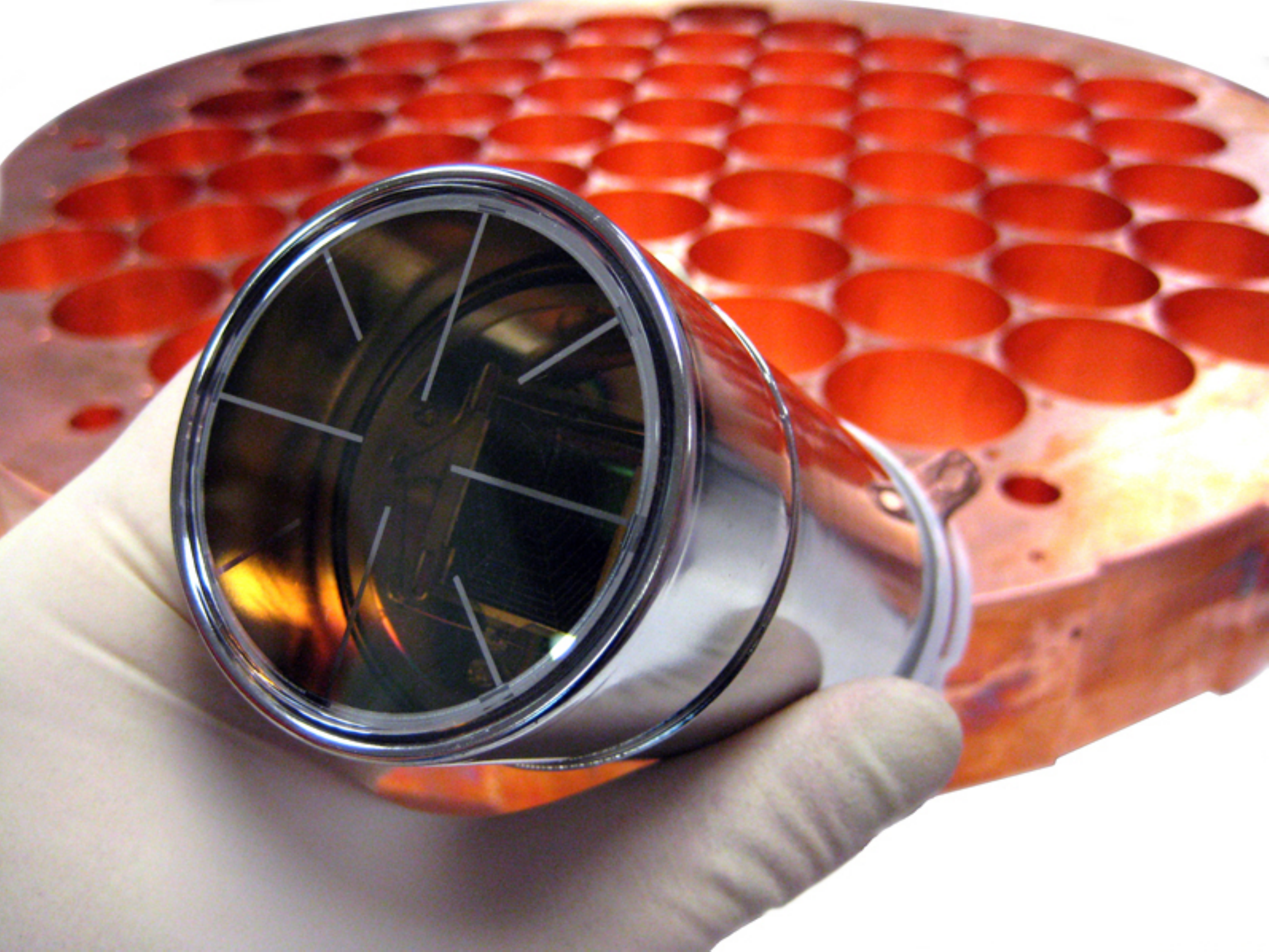}
\caption{A Hamamatsu R8778 PMT and one of the supporting copper blocks.}
\label{PMTactivity}
\end{figure}

The slow control database is used to manage all aspects of PMT handling and characterization. In this way, we can correlate changes in the PMT integrity with particular actions that were performed. Gain calibrations, AP health monitoring and other tests results are included in this database, and are integrated into the data analysis chain.

\textit{In situ} PMT gain calibrations are performed with 12 blue (430 nm) LEDs, placed inside the detector. Six LEDs are mounted on the top PMT holder to illuminate the bottom PMTs and six LEDs are installed on the bottom PMT holder to illuminate the top PMTs. The LEDs have PTFE diffusers to ensure a more uniform illumination of the PMT arrays. The LED system is also used for weekly AP measurements for PMT health monitoring.

\subsection{PMT Radioactivity}
A total of 15 LUX PMTs were counted for radioactivity at SOLO, the SOudan LOw background counting facility at the Soudan Underground laboratory in Minnesota \cite{SOLO}. The counting was performed using the Diode-M, a 0.6-kg high-purity Ge detector, inside a Pb shield, providing $>$30~cm shielding in all directions, with an 5-cm thick inner liner of low-background Pb. The shield cavity can comfortably accomodate 5 R8778 PMTs at a time. Table \ref{PMTbackground} summarizes the results obtained from these measurements.  The impact of the PMT radioactivity on the LUX background counting rates are discussed in detail in Sec.~\ref{sec:bg}.

\begin{table}
\begin{centering} 
\begin{tabular}
{|   c   |   c   |  c  |  c  |}
\hline			
U & Th & Co-60 & K-40 \\
\hline			
9.8 $\pm$ 0.7 & 2.3 $\pm$ 0.5 & 2.2 $\pm$ 0.4 & 65 $\pm$ 2 \\
\hline
\end{tabular}
\caption{Radioactivity levels of the LUX R8778 PMT, based on a 15 PMT sample. Units are mBq/PMT}
\label{PMTbackground}
\par\end{centering}
\end{table}

\subsection{PMT Health Monitoring}
\label{PMTHealthMonitoring}

The characterization of afterpulsing populations for the R8778 allows us to track the levels of residual gases inside the PMT. This is of vital importance for tracking the health of each PMT at different handling stages. The following list describes the main afterpulsing populations and the most common cause for a residual gas population to cross the PMT vacuum seal:

\begin{itemize}
\item{Elevated He$^{+}$ AP levels indicate environmental He contamination which implies poor PMT storage conditions.  Since He diffuses easily through the PMT quartz window, it is a PMT health hazard.}
\item{ Elevated N$^{+}$/O$^{+}$ AP levels are indicative of a compromised PMT vacuum seal, which allows some air to leak into the PMT interior. Assuming the PMT seals were good at the time of fabrication, this elevated AP indicates that the PMT may have experienced mechanical shock due to poor handling. PMTs with such vacuum breach should not be used in LXe.}
\item{Any sign of a Xe$^{+}$ AP level when the PMT is placed in LXe indicates a vacuum breach and an impending PMT failure. That PMT must not be used and the unit must be replaced when possible. A Xe$^{+}$ level with low N$^{+}$/O$^{+}$ levels indicates that the vacuum breach occurred inside the chamber, very likely due to stress from differential contraction in the PMT body during cooling.}
\end{itemize}

During dark matter operation, weekly AP measurements will be performed to track small changes in the PMT vacuum along with LED gain calibrations.

\subsection{Light Collection}

The efficiency of the LUX detector to collect scintillation photons has been optimized by careful positioning of PTFE reflectors around the active volume of the detector. High UV-reflective PTFE panels are used to surround the active region, a technique which has been shown to increase the light yield in previous detectors by up to a factor of 5. LUX photon collection is augmented beyond previous designs with the addition of PTFE reflectors, completely enclosing the gas phase, and a network of PTFE ``trifoils", completely filling the gaps between the PMTs. This design ensures that the active region is surrounded in 4$\pi$ by high-reflectivity material or PMTs.

Electric field grids have been designed to minimize their optical footprint in the detector. The LUX grids are 96-99\% transparent at 0$^\circ$ angle of incidence, due to thin wire diameter, large spacing, and the use of a strung pattern instead of a mesh. The exception is the anode grid, which uses a mesh design and is 88\% transparent. The grids are constructed from stainless steel, a material shown to be $\sim$57\% reflective at xenon scintillation wavelengths \cite{Bricola2007}. Further details on the grids can be found in Sec.~\ref{sec:grids}.

The LUX light collection efficiency was measured during the above-ground commissioning run, described in detail in Ref.~\cite{AkeribRun2}.  Using the photo peak from the 662~keV $\gamma$-ray from $^{137}$Cs, the light collection efficiency was determined to be 8~phe/keV$_{\text{ee}}$. Using Monte Carlo simulations, it was determined that the reflectivity of the PTFE panels is $>$~95\% in LXe and that the photo absorption length of the scintillation light in the liquid is at least 5~m \cite{AkeribRun2}.

\section{Electronics and Readout}
\label{sec:electronics}

\subsection{PMT Bias Distribution and Cabling}

Custom Cirlex$^{\tiny{\textregistered}}$ circuit boards were built to supply bias voltage to the PMTs.
Both bias and signal voltages are transmitted through custom Gore coaxial cable
with stainless steel braid (to minimize conductive heat loss) and Fluorinated Ethylene Propylene (FEP) dielectric.  The connections to the circuit boards are made with custom Cirlex$^{\tiny{\textregistered}}$
strain-relieving connectors.  The top of the circuit boards are covered with
a Teflon cap to prevent an electric short or breakdown.  All cables are heat-sunk 
and strain-relieved at the top of the copper radiation shield, just above the top 
array of PMTs.

Signal propagation in the cables is equivalent to that of RG178 cables and thus
transmit the fast photoelectron signals with minimal distortion and low loss.
The 10~m internal cable lengths result in an approximately 25\% reduction in
signal pulse height.

\subsection{Analog Electronics}

The goal of the LUX electronics is to have 95\% of the single photoelectrons in any PMT be clearly resolved from a 5$\sigma$ fluctuation in the baseline noise. The measured noise at the input of the LUX data acquisition system (DAQ) is 155~$\mu$V$_{RMS}$ (1.3 ADC counts). Since the typical resolution of single photoelectron distributions measured with the LUX PMTs is 37\%, the gain of the analog chain must put the peak of the single photoelectron distribution at 30~ADC counts.  The LUX PMTs operate with a gain of 3.3$\times$10$^6$ and a single photoelectron produces a pulse with an area of 13.3~mVns and a FWHM of 7.7~ns at the PMT base with 50~$\Omega$ termination.

Two stages of amplification are provided by the analog chain.  The preamplifier, discussed in Sec.~\ref{sec:preamplifier}, provides an effective gain of $\times$5 into 50~$\Omega$.  The postamplifier, discussed in Sec.~\ref{sec:postamp}, provides three differently shaped outputs for the DAQ digitizers, the DDC-8 digitizers of the trigger system, and the CAEN discriminators. 

\subsubsection{Preamplifier}
\label{sec:preamplifier}

The preamplifiers employed in LUX have a simple single IC (AD8099) design that provides a gain element at the interface between the xenon space of the detector and the air space of the outside world.  This avoids reflection problems due to impedance mismatch and reduces the effects of electromagnetic interference in the air space.  The nominal voltage gain of $\times$ 5 can be reduced by a factor of 10 using a selectable voltage divider at the input.  This selection is performed via an analog transmission gate in the front-end that is activated by a reduction in the supply voltage.  Supply voltages are provided by the postamplifier (see Sec.~\ref{sec:postamp}.)  

The PMT signals exit the xenon space in groups of 32.  Each preamplifier board hosts eight preamplifier channels and four preamplifier boards are housed in a single shielded enclosure.

\subsubsection{Postamplifier}
\label{sec:postamp}

The output signals from the preamplifiers are sent through 50~$\Omega$ coaxial cables to the postamplifiers, located in the detector electronics racks.
Each input channel of the postamplifier produces three output signals: the `Struck' output drives a Struck flash ADC, the `fast' output drives a CAEN V814 discriminator, and the `DDC-8' output feeds into an 8-channel analog-sum circuit, which in turn drives a DDC-8DSP digital signal processor.

The output pulse heights for a single photoelectron signal,
assuming a PMT gain of 3.3$\times10^6$, are
approximately 80~mV for the fast output,
and 5~mV for the Struck and DDC-8 outputs.
The pulse shape of the fast output is close to single-pole with the $-6$~dB 
bandwidth of 100~MHz.
The Struck and DDC-8 outputs use 4-pole Bessel filters with
cut-off frequencies at 60~MHz and 30~MHz, respectively.
Circuit simulations predict 
the input equivalent noise to be less than 2~nV/$\sqrt{\mathrm{Hz}}$,
which is dominated by the termination resistors at the PMT and at the preamplifier
input, and by the opamp in the preamplifier.
The predicted output noise levels are 0.17~mV$_{RMS}$ for the `Struck'  and `DDC-8' outputs,
and 1.8~mV$_{RMS}$ for the `fast' output.
The output dynamic range is greater than 2~V for the `Struck'  and `DDC-8' outputs.
The channel-to-channel cross talk inside the postamplifier has been measured to be
0.2\% for the `fast' output, and $<$0.2\% for the `Struck'  and `DDC-8' outputs.

The postamplifiers are implemented in 9U$\times$160~mm Eurocard modules,
with each module containing 8 channels.
The front panel carries 8 input (SMA) and 24 output (LEMO) coaxial connectors.
The DC power is supplied via a custom backplane.
A 4-pin mini-DIN connector on the front panel sends the power to the preamplifier.
A front-panel switch selects the positive power voltage for the preamplifier to be +4~V or +8~V, which selects a preamplifier gain of $\times5$
or $\times0.5$, respectively.
The preamplifier power cable is also used to send test pulses to the preamplifiers.

\subsection{Trigger Electronics}

The LUX trigger system utilizes the pulse shape information of the PMT signals to select potential dark matter events and reject background events. It has the capability to differentiate between S1 and S2 signals. Trigger decisions can be made based on just S1 signal information, just S2 signal information, or the combination of these two. Pattern recognition can be used to select events of interest based on the geometrical information provided by the PMTs. Background events, associated with large S1 and/or S2 signals and/or invalid geometrical patterns, can be vetoed. Trigger decisions are made within a couple microseconds of a valid S2 signal. 

\begin{figure}[]
\centering
		\includegraphics[width=\columnwidth]{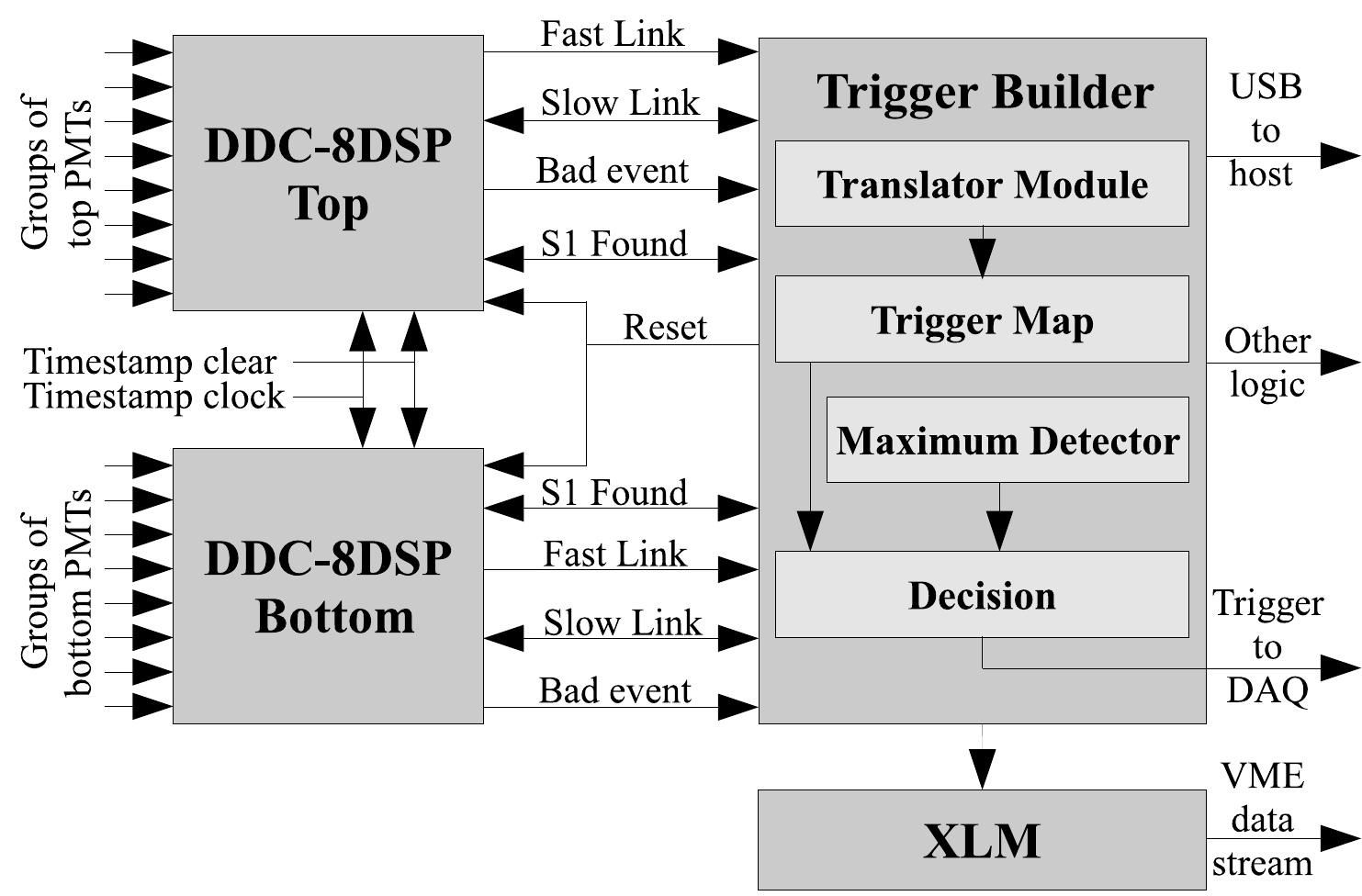}
\caption[BoardComm]{Schematic of the communication between the components of the LUX trigger system}
\label{fig:BoardComm}
\end{figure}

The operation of the LUX trigger is shown schematically in Fig.~\ref{fig:BoardComm}.  The trigger system uses two 14-bit, 64 MHz, 8-channel digital signals processors (DDC-8DSP).  Information about the detection of S1 and S2 signals is shared with the Trigger Builder (TB) where the trigger decision is made.  In order to utilize all 16 channels of the signal processors, the top and bottom PMTs have been summed into 16 groups, 8 for the top PMTs and 8 for the bottom PMTs.  The analog sums of the PMT signals is produced by LeCroy 628 Linear Fan-Ins. Each group is connected to a dedicated channel in one of the two DDC-8DSP boards. One DDC-8DSP module handles the groups associated with the top PMTs; the other handles the bottom PMTs.  The Trigger Builder (TB) is connected to the DDC-8DSPs, as shown in Fig.~\ref{fig:BoardComm}. The Fast and Slow Links use HDMI cables; the remaining signals use regular NIM cables. The fast unidirectional link uses four differential pairs of the HDMI cable and is used to transfer big blocks of data, such as S1 and S2 hit vectors, S1 and S2 multiplicities, waveforms, and timestamps. The slow link uses seven single ended bidirectional signals of the HDMI cable and is used for communicating the finite state machine states across the boards.  The TB takes $<$1~$\mu$s to process the hit-vectors from the DDC-8DSPs and generate the final trigger signal that is send to the DAQ. The TB has one dedicated HDMI connection to send operational, configuration, and diagnostic information to an XLM VME module \cite{JTec} which allows trigger information to be merged with the DAQ data stream.

The DDC-8DSPs and TB are controlled via a USB 2.0 interface, using a dedicated host computer. The host PC control software is written in BlackBox Component Builder (developed by Oberon microsystems \cite{BlackBoxOberon}) and is the component-based development environment for the programming language Component Pascal. Combined with LibUSB, an open-source USB driver \cite{LibUsb}, and DevC++ \cite{DevCpp}, BlackBox has become a successful tool for controlling the trigger system.

The LUX trigger system can be operated in three trigger modes: S1Mode, S2Mode, and S1\&S2Mode.  S1Mode and S2Mode are very alike, except that they search for different pulse types: S1 pulses and S2 pulses, respectively.  These trigger modes require a quiet period for all channels before an S1 or S2 is detected.  During a 0.1-8~$\mu$s period after an S1 or S2 signal is detected, the DDC-8DSP looks for signals in the other channels.  The hit information is stored in S1 and S2 hit vectors.  S1\&S2Mode is a combined trigger mode which requires the detection of  at least one S1 and one S2 pulses. The minimum and maximum time between the detection of the S1 and S2 pulses allows us to make event selections based on event depth. 

All trigger modes utilize a trigger map which is used to select events based on the location of the groups of PMTs in which signals are observed. The trigger map and other parameters, such as the upper and lower thresholds, the length of the quiet period, and the minimum and maximum time between S1 and S2 signals, can be adjusted by the user via the BlackBox interface, or settings in the DAQ GUI.

\subsection{Data Acquisition}
\label{sec:daq}

The DAQ system \cite{Akerib2012DAQ} uses 16 Struck 8-channel Fast ADC modules, model number
SIS3301. These ADCs digitize at $100\,\mbox{MHz}$ with a resolution
of 14 bits. They have $2\times128\,\mbox{k}\,\mbox{sample}$ dual
memory banks that allow for acquisition on one bank while data is
being downloaded from the other.  Each board is connected
to a VME bus, which is connected to the DAQ computer via a fiber
optic Gbit connection (Struck communication modules SIS3100 and SIS1100).
The boards are capable of downloading to the acquisition computer
at the maximum VME download speed of $80\,\mbox{MB/s}$ (2eVME protocol).
Each board is controlled by four FPGAs (one per pair of channels)
whose firmware was developed by Struck in collaboration with the LUX collaboration to operate in Pulse Only Digitization (POD) mode,
described in detail in Ref.~\cite{Akerib2012DAQ}. The Struck inputs have been fitted
with an on-board single-pole anti-aliasing filter of $30\,\mbox{MHz}$
on each channel. A detailed description of the DAQ system and its performance can be found in Ref.~\cite{Akerib2012DAQ}.

\section{Water Shield and Muon Tagging System}
\label{sec:watershield}

\subsection{Water Tank}

The LUX cryostat is shielded from background radiation with a 7.6-m diameter, 6.1-m high water tank, shown in Fig.~\ref{lux-ug-Fig}.  Compared to standard lead and polyethylene shields, a water shield achieves much lower $\gamma$-ray backgrounds and provides superior shielding of neutrons from cavern radioactivity and the very high-energy tail of neutrons from muon interactions in cavern walls.  The water shield is very cost competitive, particularly as the detector size increases, and provides great flexibility in modifying the size and shape of the central detector.

The water shield provides a minimum shielding thickness of 2.75~m at the
detector top, 3.5~m at the sides, and 1.2~m below.  Backgrounds associated with cavern sources of $\gamma$ and neutron activities are reduced to a level where the internal radioactivity from detector components is the dominant source of background (see Sec.~\ref{sec:bg}).

The LUX detector is lowered into the tank through a 1.2~m diameter opening at the top of the tank. Once in the tank, it is fastened to a metallic frame, anchored inside the tank, which provides stability and safety. The tank is hermetically sealed once the detector is installed and nitrogen gas is circulated continuously above the surface of the water in order to prevent radon accumulation. The water can be circulated through an industrial purifying system, designed by South Coast Water Co., in order to maintain appropriate levels of purity. The maximum acceptable level of U/Th/K impurities in the water is 2~ppt/3~ppt/4~ppb, which is about 6 orders of magnitude lower than their concentration in the surrounding rock.  These purity levels are readily achievable with a commercial purifier.

The detector is leveled using three leveling rods that can be moved up or down. The end of these rods are connected to a leveling ring that is attached to the top of the detector. The width of the S2 pulse increases as the gap between the
anode grid and the liquid surface increases. In an unleveled
detector, the S2 width will be largest on the highest side of the detector and
smallest on the lowest side. The vertical positions of the leveling rods are adjusted until the width of the S2 pulse is uniform across the detector, independent of the horizontal interaction position.  After adjusting the
leveling rods, the detector can be leveled to within $\pm$~250 $\mu$m of horizontal.  Three parallel-plate style level sensors are also used to measure the tilt of the detector, independent of the S2 response of the detector.

Aside from the central 1.2~m diameter central opening, the tank is equipped with several secondary access ports. Two 1.8~m diameter ports are located on each side to the east and west, and can potentially be used by other experiments to take advantage of the water shielding. Two additional 0.6~m diameter ports are used for maintenance and equipment feedthroughs. There is also one manhole hatch on the top surface, and one on the side of tank at about ground level.

The walls, floor, and ceiling of the tank are lined with Tyvek reflectors in order to optimize the collection of Cherenkov light emitted by passing muons.

\subsection{Muon Tagging System}

The water tank is instrumented with 20 ten inch PMTs, and incoming muons can be tagged as they enter the water and emit Cherenkov radiation.  Any nuclear recoil events in the LUX fiducial volume that are coincident with a muon in the water tank are eliminated during data analysis. A single downward-going muon will, on average, produce about 200 photoelectrons.  The amount of light collected from muons that produce a hadronic shower in the tank will be even greater.  Thus, essentially all showering muons passing within two meters of the edge of the detector can be vetoed.  It should be noted that the 2.4-meter thick reflective veto in Super-Kamiokande \cite{SuperK} is estimated to be more than 99.99\% efficient with similar photomultiplier coverage.

\begin{figure}[h]\begin{center}
\scalebox{0.75}{\includegraphics[width=\columnwidth]{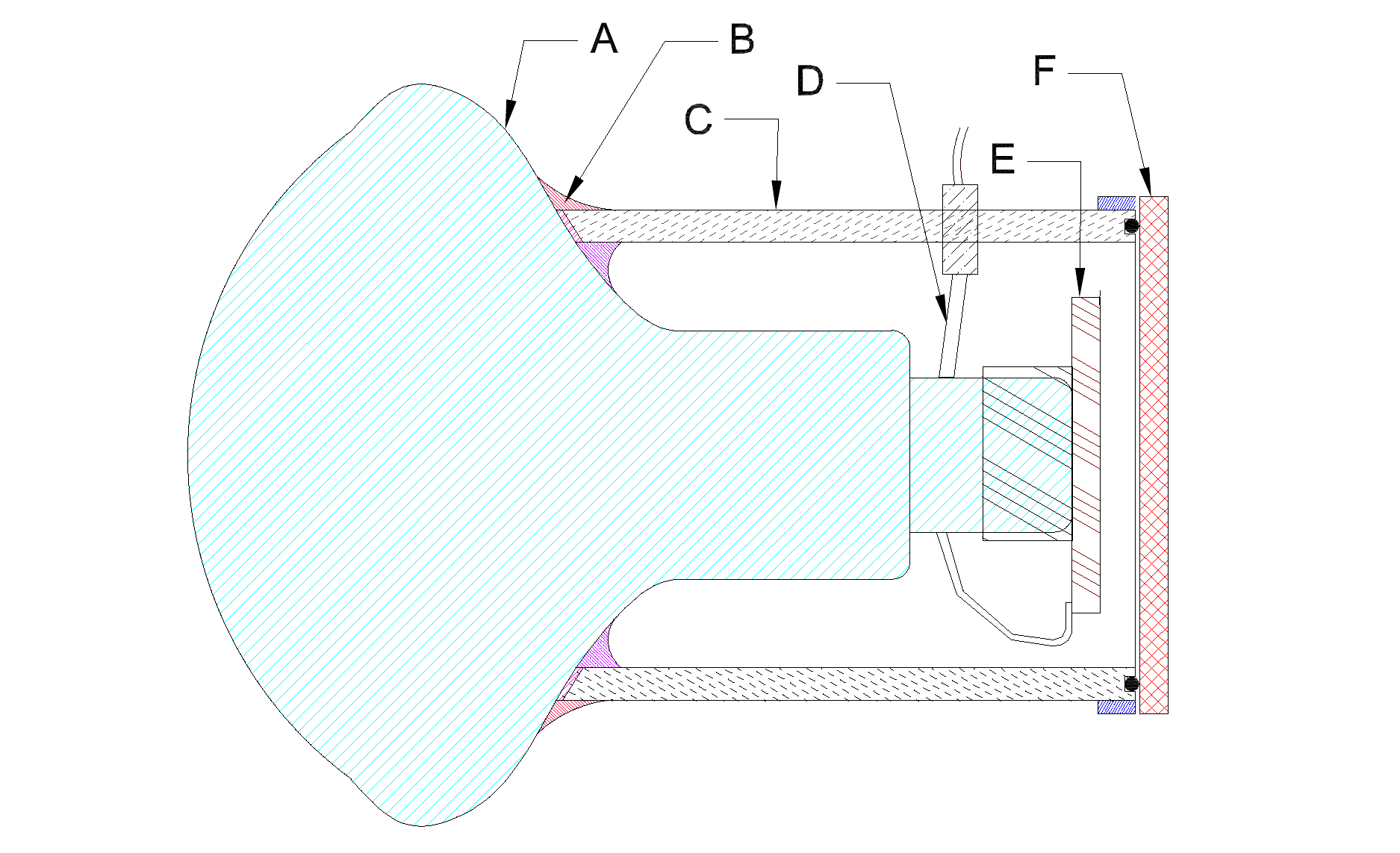}}
\caption{The water-tight housing for the water tank PMTs.  The labels correspond to the following: A. Hamamatsu R7081 10-inch diameter PMT, B. Adhesives, C. 6 inch schedule 80 PVC pipe, D. RG-58P coax cable, E. PMT bias board, and F. clear acrrylic end-plate.}
\label{fig:pmt_housing}
\end{center}\end{figure}

The PMTs used for the water tank are Hamamatsu R7081 PMTs. The PMTs are placed in waterproof housings and use positive high voltage.  Each PMT is epoxied to a PVC pipe and sealed with a removable clear acrylic back.  An RG-58 cable is fed through the side of the PVC pipe and connected to a feed-through at the top of the water tank.  The cable jacket is water-blocked to prevent water from entering the housing. The PMT housing is weighed down to counter-balance the buoyant force of the PMT cathode and waterproof housing.  Figure~\ref{fig:pmt_housing} shows a sketch of the PMT with its watertight housing.

The PMTs are arranged in four vertical strings, one foot from the edge of the tank.  Each string has four PMTs, with one additional PMT located at the bottom of the tank, two feet away from the edge.  The PMTs on the string are mounted on two parallel cables, held in place by eyebolts on the top and bottom of the tank.  The remaining PMTs that sit on the bottom of the tank are weighed down with PVC piping attached to the PMT housing and filled with iron pellets.  The RG-58 cables inside the water tank were chosen for their low-radon outgassing properties. The cables connect to a break-out box on the top of the tank.  From there, standard high-voltage cables are used to reach the electronics.

Because the PMTs used in the system are supplied with positive high voltage, they require a circuit to capacitively pickoff the fast signal pulses. The parts were specially chosen to provide a good high-frequency response, especially capacitors with a COG dielectric, and rated up to 3~kV.  The amplifiers used provide two identical outputs; one output is digitized by the Struck ADCs (see Sec.~\ref{sec:daq}) while the other signal is sent to a discriminator and a fast trigger system composed of off-the-shelf NIM modules to generate a muon trigger.  Muon trigger data will be recorded to monitor backgrounds observed in the water tank.  The information provided by the muon tagging system will be used to veto events off-line in which activity in the water tank was recorded in coincidence with signals in the fiducial volume of the central detector.

\section{Background Studies}
\label{sec:bg}

The LUX background model and screening program were constructed with
the goal of maintaining a background expectation of $<$1~WIMP-like
background event in 30,000~kg~days. This goal lowers the WIMP sensitivity
limit of the experiment below 2$\times$10$^{-46}$~cm$^2$
for a WIMP mass around 40-50~GeV/c$^2$, nearly an order of magnitude
lower than the lowest limit from current dark matter experiments at the same mass \cite{Aprile2012}.
This sensitivity corresponds to a 14~$\mu$DRU$_\text{nr}$ event rate within the fiducial volume over the energy
range 5-25~keV$_\text{r}$.
This rate corresponds to an upper limit of 8.6~background
events within the 100~kg fiducial volume for a 300~day WIMP
search, reduced to $<$~3.9~events at 90\% CL after applying the 45\% nuclear recoil
acceptance cut.

LUX background levels are greatly suppressed by the use of an unprecedented
amount of liquid xenon. The self-shielding properties of liquid xenon,
combined with fiducialization and multiple-scatter event rejection,
serve to reject WIMP-like $\gamma$-ray backgrounds for LUX at the level of $\sim$10$^{-6}$,
before consideration of recoil band discrimination factors, typically
at the level of $>$99.5\% for liquid xenon. Nuclear recoil backgrounds
are reduced at the level of $\sim$10$^{-3}$, more than sufficient to
render neutron backgrounds subdominant to $\gamma$-ray contributions.

An extensive background modeling campaign using the LUXSim package \cite{Akerib2012LuxSim}
has been performed in order to assess the impact of both internal
and external backgrounds. For estimates of background contributions
from internal components, $\gamma$ simulations use energy spectra from
all major isotopes and their decay chains, measured for each internal
component. Neutron simulations combine calculated ($\alpha$,n) and
fission spectra for measured levels of $^{238}$U and $^{232}$Th.

\subsection{Internal Backgrounds}

Internal contributions are expected primarily to come from the PMT
arrays, due to their proximity to the active region and variety of
materials used in their construction. Internal background benchmarks
are set by measurements of PMT radioactivities listed in Table~\ref{PMTbackground}. Upper limit measurements from Hamamatsu were initially
used to generate conservative background benchmarks; subsequent counting
at SOLO found the PMTs to have overall $\gamma$-ray activities reduced by
over a factor of $\times$1/2. LUXSim modeling using current SOLO
measurements predict a single-scatter fiducial $\gamma$-ray event rate of
437$\pm$37~$\mu$DRU$_{\text{ee}}$ in the energy range 1.3-8~keV$_{\text{ee}}$,
corresponding to 88$\pm$7~events within the 100~kg fiducial volume
in 300~days, before applying electron-recoil 
rejection cuts.

The neutron production rate of the Hamamatsu R8778 PMTs used by LUX
has not been directly measured, but a reasonable estimate can be made.
Neutrons are generated in the PMT through fission and ($\alpha$,n)
processes. The ($\alpha$,n) neutron production rate is a function
of the chemical composition of the target material for the $\alpha$-particles,
i.e. the components in the vicinity of the decays site. The chemical
composition of the R8778 PMT internal components has been made available
by Hamamatsu to the LUX Collaboration \cite{Hamamatsu2007}, but is
protected by a confidentiality agreement. The mass of the PMT is dominated
by Fe, Ni, and Co in the steel body and the electrodes, and by O
and Si in the glass of the window and stem. The ($\alpha$,n)
neutron production rate for each one of the listed compounds can be
calculated using the ``Neutron Yield'' tool \cite{neutronyield}
based on the techniques described in Ref.~\cite{Mei2009}. Assuming that
the aforementioned materials are evenly spread throughout the PMT
volume, and using the PMT contamination levels measured at SOLO, the
neutron production rate combined for fission and ($\alpha$,n) processes
is calculated to be $1.2$~n/PMT/yr \cite{Akerib2012PMT}.  Recent results obtained by the XMASS collaboration (oral presentations only at the time of this writing), using the same R8778 PMTs as LUX, hint at a possible break in equilibrium in the $^{238}$U chain and a high concentration of radioactive impurities in the small aluminum ring around the perimeter of the PMT faces. LUX performed simulations of the worst-case scenario, assuming that all of a PMT's radioactivity originates from the Al ring, with the upper limit levels indicated by XMASS. The result is a total neutron background rate in the fiducial volume which is a factor $\times$5 above the baseline scenario described above.

To estimate the background contributions from each major internal
component, LUXSim Monte Carlo estimates of $\gamma$-ray and neutron background
contributions were generated using a faithful recreation of the precise
LUX geometry, including all internals to CAD specifications, as well
as the water shield and outer components. Gamma simulations use energy
spectra from all major isotopes and their decay chains, measured for
each internal component, while neutron simulations combine calculated
($\alpha$,n) and fission spectra for measured levels of $^{238}$U
and $^{232}$Th. These simulations yield a scaling factor in units
of DRU$_{\text{ee}}$/mBq or DRU$_{\text{nr}}$/mBq for each major
isotope identified in the counting process. Counting results for major LUX internals, 
described in Sec.~\ref{sec:screening}, are combined with the Monte
Carlo background estimates to predict the total electron- and nuclear-recoil contributions
from all components. Totals are listed in Table \ref{tab:ER-NR-evt-rates}.

\begin{table}
\centering
\begin{tabular*}{1\columnwidth}{@{\extracolsep{\fill}}|>{\centering}p{0.2\columnwidth}|>{\centering}p{0.2\columnwidth}|>{\centering}p{0.2\columnwidth}|}
\hline 
& \multicolumn{2}{c|}{WIMP-Like Events}\tabularnewline 
& \multicolumn{2}{c|}{30,000 kg~days)}\tabularnewline
\hline 
 & ER & NR\tabularnewline
\hline 
\hline 
PMTs & 0.4 & 0.03\tabularnewline
\hline 
Cryostats & $<$0.02 & $<$0.002\tabularnewline
\hline 
Grid wires & $<$0.01 & $<$0.001\tabularnewline
\hline 
PTFE panels & $<$0.05 & $<$0.009\tabularnewline
\hline 
HDPE & $<$0.01 & $<$0.002\tabularnewline
\hline 
Cu & 0.03 & $<$10$^{-4}$\tabularnewline
\hline 
$^{85}$Kr & $<$0.13 & --\tabularnewline
\hline 
\end{tabular*}

\caption{ER and NR projected activity contributions
from all major internal components in the energy range between 1.3 and 8~keV$_{\text{ee}}$. Measured material activities from
Table \ref{PMTbackground} are combined with Monte Carlo simulations
to predict the total background contribution from each individual
component. The ER event rate from the copper components is estimated
from \cite{laubenstein2009}, pending a final measurement of copper
samples exposed at the Sanford Surface Laboratory for the duration
of LUX above-ground running. The $^{85}$Kr event rate is estimated
from the projected efficiency of the LUX Kr removal system.}
\label{tab:ER-NR-evt-rates}
\end{table}

Over 200 materials were used during the LUX construction phase. To
estimate the background contribution from each of these components,
a generic Monte Carlo background "map" was created for 15 separate
regions in the detector. DRU/mBq scaling factors were obtained for
each region and each particular isotope. Minor components then inherit
the scaling factor from their corresponding regions, and screening
results are integrated to obtain a total contribution. For ($\alpha$,n)
background estimates, the dominant surrounding material in each region
is used. The screening program for minor components is currently ongoing.

Particular attention has been paid to modeling the multiple-scintillation single-ionization (MSSI) 
background, in which a $\gamma$-ray undergoes multiple scattering with one
of the scattering vertices in the field region and the rest outside
of the active region, e.g. in the reverse-field region (RFR) below
the cathode grid. As only the scattering vertex within the field region
will generate charge for an S2 signal, but S1 light from all vertices
will be counted as a single event, the relative S2/S1 size will be
shifted downward, greatly reducing the ER rejection efficiency for
these events. The goal is to reduce the fraction of MSSI events
to total $\gamma$-ray events below the level of (1~-~ER~rejection), or
6$\times$10$^{-3}$ assuming a very conservative 99.4\% ER discrimination. Monte Carlo studies of these event types were performed in order to
optimize the detector geometry for minimization of MSSI events.
Using the expected dominant PMT $\gamma$-ray model, it is predicted that
a MSSI fractional level below $6\times10^{-3}$ will be obtained
with a RFR height $\le$5~cm for events depositing 1-50~keV$_{\text{ee}}$
in the fiducial volume; for the LUX RFR height of 3.5~cm, a fractional
level of $4\times10^{-3}$ is projected. Note that this does not include
additional rejection of MSSI events through analysis cuts.

\subsection{External Backgrounds}

\begin{figure}
\centering
\includegraphics[width=0.9\columnwidth]{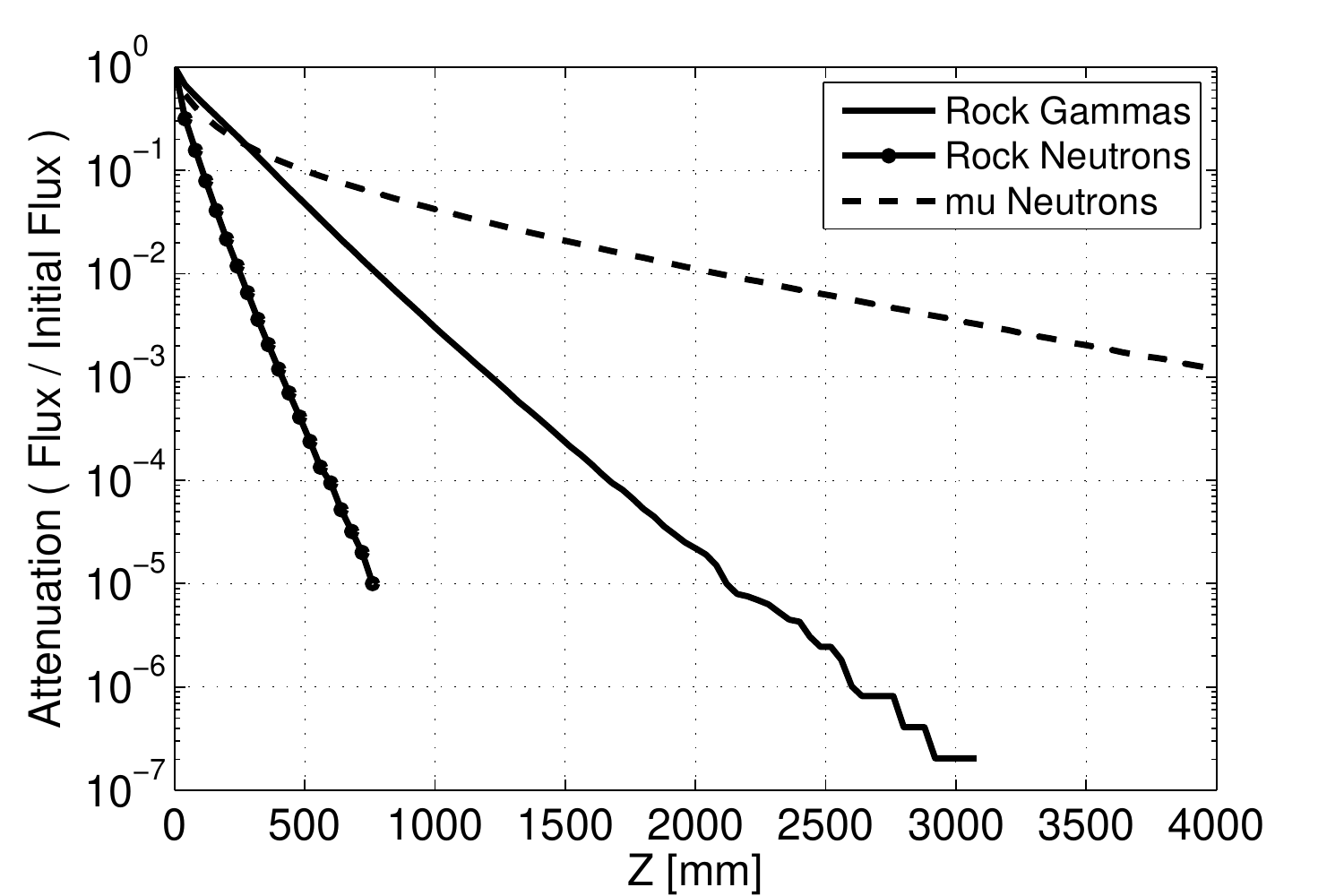}
\caption{Measured $\gamma$-ray and neutron flux reduction
in water. The flux reduction is measured in a semi-infinite wall of
water, and is defined as the number of particles that travel beyond
a depth Z, divided by the number of emitted particles. For the neutron
flux, an energy cut of E$_n$~$>$~1~keV is applied to remove low-energy
and thermal neutrons.}
\label{fig:Gamma-neutron-flux-water}
\end{figure}

External neutron backgrounds are expected to be rendered subdominant
to internal backgrounds due to the use of the water shield described in Sec.~\ref{sec:watershield}.
The water shield size
was chosen primarily to bring the high-energy muon-induced neutron
flux well below the background target. Water shield moderation efficiency
for various external backgrounds is shown in Fig.~\ref{fig:Gamma-neutron-flux-water}.

External $\gamma$-ray backgrounds will be reduced well below the internal
background by the water shield and the steel pyramid. The $\gamma$-ray flux from
the cavern rock will be reduced by a factor of $8\times$10$^{-9}$
through water shielding alone, with a further factor of $\times$1/40
reduction from the steel configuration. This is expected to reduce
the external $\gamma$-ray  background contribution to $<$6$\times$10$^{-3}~\gamma$
in the $1.3-8$~keV$_{\text{ee}}$ energy range in 30,000 kg$\cdot$days,
assuming a highly radioactive cavern composed entirely of rhyolite, producing
$\sim9~\gamma$/cm$^2$/s 
at the water shield outer edge.

External neutron backgrounds are generated primarily from high-energy
muons interacting in both the cavern rock and the water shield. At
the 4850~level of the Homestake mine, shielding from muons is provided
by a 4,300~m water-equivalent rock overburden, reducing the muon flux to 4.4$\times$10$^{-9}$$\mu$/cm$^2$/s.
There are no measurements of the $\mu$-induced neutron flux in the
Davis laboratory. However, the neutron flux from the rock has been
measured at several underground sites, and the neutrons flux vs. depth
can be fitted by an empirical equation \cite{Mei2006}. The neutron
flux from the rock at Homestake is calculated to be $(0.54\pm0.01)\times10^{-9}$~n/cm$^2$/s,
corresponding to an event rate of 163~nDRU$_{nr}$.  Requiring a single-scattering in the fiducial volume reduces the background rate by a factor of 3, resulting in a rate of 54~nDRU$_{nr}$ 
in the  $5~-~25$~keV$_r$ energy range. This corresponds to a total
event rate of $\sim0.03$ NR events in 30,000 kg$\cdot$days
fiducial.

The background contribution from neutrons produced in the water shield is expected to double that of rock-produced neutrons.
The total activity for water-produced neutrons is calculated to be
121$\pm$8~nDRU$_{nr}$ after application of single-scatter and fiducial
cuts, corresponding to 0.07~events in 30,000~kg$\cdot$days fiducial.
Event rates corresponding to these neutron backgrounds can be found
in Table \ref{tab:LUX-event-rates}. The effectiveness of the water
shield against muon-induced neutrons is further enhanced by the muon
tagging system, in which the instrumented water shield will act as a Cherenkov light detector with $>$90\%
muon rejection efficiency.

Background models assume that the concrete used in the Davis laboratory
will match the radioactive contamination levels in the surrounding
rock (0.16/0.20~ppm or 2/0.8~Bq/kg for $^{238}$U/$^{232}$Th),
and no increase to the $\gamma$-ray and neutron background
is expected. Recent radioactive screening of the concrete mixes confirm that assumption as valid and conservative.  From the ratio of contamination levels, we expect the
fast neutron flux at the Davis laboratory to be $\times37$ lower
than the flux at Gran Sasso, or $16.2\times10^{-9}$~n/cm$^{2}$/s
for E$_n>1$~MeV \cite{Mei2006}. The fast neutron background is moderated very
efficiently by the water shield. The flux is attenuated
by a factor of $3\times10^{-4}$ in the first 50~cm of water, as is shown in 
Fig.~\ref{fig:Gamma-neutron-flux-water}). The fast neutron flux is
attenuated by a factor of $2\times10^{-15}$ by 2.5~m of water, and
by a factor of $5\times10^{-21}$ by 3.5~m of water. The integrated
flux is reduced by $6\times10^{-22}$, corresponding to only $\sim2\times10^{-16}$
neutrons/year incident on the cryostat surface.

\begin{table}[t]
\centering
\begin{tabular}{|>{\centering}m{0.3\columnwidth}|>{\centering}p{0.3\columnwidth}|>{\centering}p{0.3\columnwidth}|}
\cline{2-3} 
\multicolumn{1}{>{\centering}m{0.3\columnwidth}|}{} & Differential Event Rate  & Event Total\tabularnewline
\multicolumn{1}{>{\centering}m{0.3\columnwidth}|}{} & fiducial volume (100kg) & 30,000~kg-days (fiducial)\tabularnewline
\multicolumn{1}{>{\centering}m{0.3\columnwidth}|}{} & $5-25$ keV$_{r}$ ($1.3-8$ keV$_{\text{ee}}$) & $5-25$~keV$_r$ ($1.3-8$ keV$_{\text{ee}}$)\tabularnewline
\hline 
$\gamma$'s from rock & $0.54$~ndru$_{\text{ee}}$ (no rhyolite) & $1\times10^{-4}$\tabularnewline
\cline{2-3} 
 & $27$~ndru$_{\text{ee}}$ (all rhyolite) & $6\times10^{-3}$\tabularnewline
\hline 
$\mu$-induced high energy neutrons from water shield & $\left(121\pm8\right)$~ndru$_r$ & $(73\pm5)\times10^{-3}$\tabularnewline
\hline 
$\mu$-induced high energy neutrons from rock & $\sim54$~ndru$_r$ & $\sim33\times10^{-3}$ \tabularnewline
\hline 
\end{tabular}

\caption{LUX event rates for external
backgrounds, using the full LUX shield.}
\label{tab:LUX-event-rates}
\end{table}

\subsection{Materials Screening Program}
\label{sec:screening}

The components that are used for the LUX detector are $\gamma$ screened to ensure that the
overall background goals are met. The SOLO germanium detector at the Soudan Mine is being used for the screening program.  SOLO is able to reach
sensitivities of 0.1 ppb U, 0.1 ppb Th, 0.25 ppm K. The screening facility can accept large 
samples ($8"$ $\times$ $8"$ $\times$ $8"$) which
allows it to be used for batch screening of the PMTs.

LUX has also access to the LBL counting facility at Oroville. It is through this program that a source of ultra-low radioactivity titanium was identified for the cryostat \cite{Akerib2012Ti}. The majority of all relevant parts of the detector's internals are organized in a database which is directly linked to a simplified LUX background model software. This provides a convenient way to assess a particular component's potential danger to the experiment's background, and easier prioritization of the counting queues.

\section{Integration and Underground Deployment}
\label{sec:integration}

\subsection{Assembly of the Main Detector}

The detector subsystems were fully integrated at the surface facility at SURF.  The TPC was assembled and sealed within the inner and outer cryostats in a clean room.  The cryostat domes were suspended from a structure mounted on rails while the TPC was assembled.  The cryostat cans were initially located in a pit within the clean room  and mounted and removed using a counter-weight system.  After sealing the TPC within the inner cryostat, cold heads, heat exchangers, and thermal sensors were mounted on the exterior of the inner cryostat.  After sealing the external cryostat, and with the flexible conduits connected between the dome and the breakout assembly, the completed detector was rolled on rails out of the clean room and positioned above the water tank.

The installation of the detector into the water tank is a step-wise process, due to the limited headroom available in the Davis Lab.  The center of mass of the detector when it is installed in the water tank is located 5.2 m below the 
deck level. The overhead space constrains the maximum hook height of the hoist to 2.7 m 
above the deck.  This vertical geometry was duplicated at the surface facility to be able to exercise the design and the methods of installation prior to deployment of the detector underground.  

The detector is removed from the cart with the hoist and mounted on the detector supports located at the deck level, while the first stage of the supporting structure is attached.  The detector is then lifted off the supports and lowered below the deck level.  This process is repeated, and the flexible conduits and rigid thermosyphon plumbing and conduit are fixed to the supporting tower structure.  When the second tower sections are installed, the detector is lowered onto the stand in the water tank and secured in place using fasteners engaged from the deck above.

Transport of the detector assembly to the Davis Cavern occurred without opening the inner TPC cryostat or any of the flexible conduits that are extensions of the xenon gas space.  The detector and breakout carts were designed to constrain the detector into a volume 11.3 m long, 1.4 m wide, and 2.6 m high.  These dimensions were imposed by the dimensions of the Yates hoist conveyance.   The cryostats were cushioned by coiled-cable springs upon which the external dome is mounted to the detector cart.  
The carts were transported from the surface laboratory to the Yates cage using a forklift. 
The rail wheels were removed and pneumatic cushions were installed underneath.  The carts were gently pushed in and out of the cage, and gently moved all the way to the doors of the Davis laboratory.  At this point, the carts was transitioned back to rail wheels. After the PMTs and grids were checked, the detector was installed into the water tank.  Commissioning of the detector started in Fall 2012.

\subsection{Below-ground Integration}

\begin{figure}
\includegraphics[width=\columnwidth]{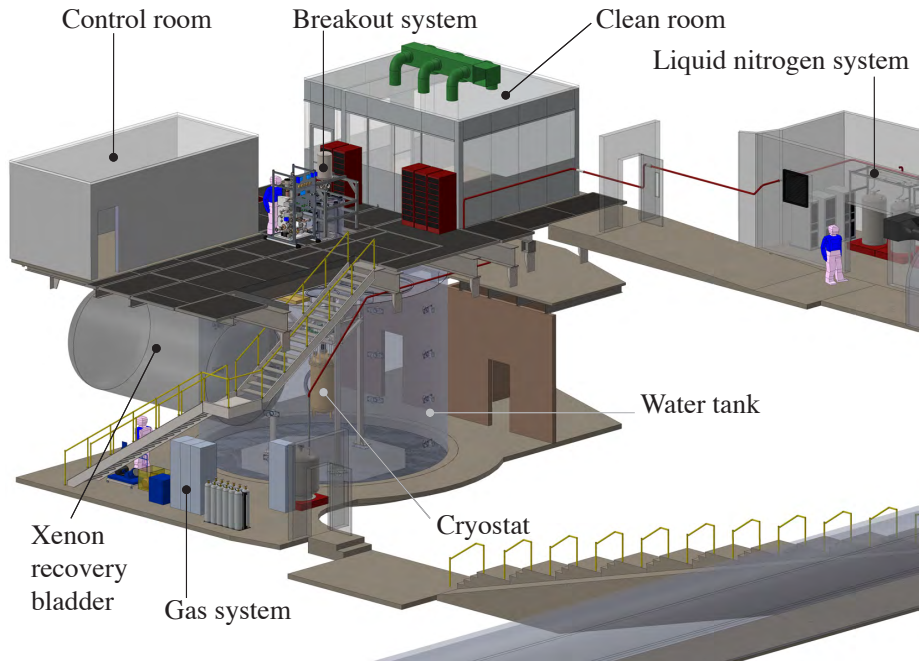}
\caption{Rendering of the layout of the Davis Laboratory.}
\label{fig:Davis_layout}
\end{figure}

In July 2012, the LUX detector was moved into the newly re-commissioned Davis Underground Laboratory. Located 4850 feet deep within the former Homestake gold mine near Lead, SD, it consists of a single cavern, $\sim$10.5~m long,  $\sim$18~m wide, and $\sim$12~m high. Between 1970 and 1994, the cavern was used by Ray Davis for his solar neutrino experiment. After it was decommissioned and the mine closed down, water filled the galleries to levels well above the Davis laboratory. It was only in May 2009 that the 4850 ft level became available again. Since then, intensive work has been taking place to clean up and renovate the laboratory, to bring it up to modern standards of safety. The LUX collaboration has been tightly involved in all phases of the re-design of the cavern. A rendering of the laboratory's main features can be found on Fig.~\ref{fig:Davis_layout}.

The LUX water tank sits on the floor of the cavern (level 0), on top of $\sim$20~ton of low-radioactivity steel acting as an additional shield used to decrease the overall $\gamma$-ray even rate in LUX by an additional factor of 40. The xenon gas system and the storage vessels are also located on this level. An emergency xenon recovery bladder hangs below the deck structure to the south of the water tank. It can expand to fill the available space there and contain all 370 kg of xenon in warm gas state, if all other recovery methods fail.

A metallic deck was constructed 7.6~m above floor level to support additional equipment and infrastructure specific to LUX. It sports a noise-insulated control room in the south-west corner, and a class 1,000 clean room on the north side for detector assembly work. The electronics and slow-control systems are also located on this level. The main entry into the laboratory is made at deck level from the south side, through a newly excavated access tunnel. The entire cavern space is maintained at a class 100,000 clean space level and isolated from the rest of the underground complex by air-tight doors. In order to enter the Davis laboratory, all personnel and equipment has to go through a clean-transition area, located in the new access tunnel; personnel go through adequate cleaning procedures and changes into clean clothing, while equipment goes through an airlock to be scrubbed and cleaned. The transition space is shared with the Majorana collaboration, who run a low-background facility adjacent to it.  The LUX detector followed that route to enter the laboratory. After entering the laboratory, the detector and the breakout carts were lowered onto a rail system, running North-South on the deck, passing over the central opening of the water tank, and continuing all the way into the clean room on the north side. During normal operations, the breakout cart sits just to the south of the central tank opening, connected through a radon-tight seal to the detector inside the tank, while the detector cart is moved away. During testing and/or detector assembly in the clean room the detector is installed on its cart, which acts as a working stand, while the breakout cart remains just outside the clean room doors. This allows us to operate most electronics and diagnostic systems through the breakout cart even while the detector is in clean room.

\section{Summary}
\label{sec:summary}
In this paper we have described the design of the LUX dark matter detector, installed at the Sanford Underground Research Laboratory in Lead, South Dakota.  The detector has been assembled and commissioned at the surface facility of SURF and was transported to the 4850 level in July 2012.  Detector commissioning started in Fall 2012 and operation is expected to start in early 2013.

\section{Acknowledgements}
This work was partially supported by the U.S. Department of Energy (DOE) under award numbers DE-FG02-08ER41549, DE-FG02-91ER40688, DOE, DE-FG02-95ER40917, DE-FG02-91ER40674, DE-FG02-11ER41738, DE-SC0006605, DE-AC02- 05CH11231, DE-AC52-07NA27344, the U.S. National Science Foundation under award numbers PHYS-0750671, PHY-0801536, PHY-1004661, PHY-1102470, PHY-1003660, the Research Corporation grant RA0350, the Center for Ultra-low Background Experiments in the Dakotas (CUBED), and the South Dakota School of Mines and Technology (SDSMT). LIP-Coimbra acknowledges funding from Funda\c{c}\~{a}o para a Ci\^{e}ncia e Tecnologia (FCT) through the project-grant CERN/FP/123610/2011. We gratefully acknowledge the logistical and technical support and the access to laboratory infrastructure provided to us by the Sanford Underground Research Facility (SURF) and its personnel at Lead, South Dakota.

\end{document}